%% file: Railways_PNW.tex
\documentclass[12pt,twoside]{article}
\usepackage{setspace}
\usepackage{amsmath}
\usepackage{bm}
\usepackage{mathtools}

\usepackage{Style_Railways}
\usepackage{adjustbox}
\usepackage[numbered]{bookmark}
\usepackage{indentfirst}
\usepackage{caption}
\usepackage[multiple]{footmisc}

\usepackage{multirow}
\usepackage{makecell}
\shortcites{bahgat2018,kalemli2015construct}
\usepackage[english]{babel}
\usepackage{natbib}

\definecolor{blue(pigment)}{rgb}{0.2, 0.2, 0.6}
\begin{document}
\bibliographystyle{chicago}

\input{Tables/Values}

\newdateformat{mydate}{\monthname[\THEMONTH]{ }\THEDAY, \THEYEAR}

%plainnat   aernobold   apalike

\begin{spacing}{1.2}
\begin{titlepage}

\title{{\LARGE Production Networks and War: \\ Evidence from Ukraine}%
\thanks{{We thank Alireza Tahbaz-Salehi, Francesco Amodio, Michal Bauer, Lori Beaman, Eli Berman, Mathieu Couttenier, Georgy Egorov, Konstantin Egorov, Alessandro Ferrari, Renata Gaineddenova, Luigi Guiso, Christian Hellwig, Stepan Jurajda, Amit Khandelwal, Alexander Libman, Claudio Michelacci, Ameet Morjaria, Dmitry Mukhin, Nathan Nunn, Lavinia Piemontese, Nancy Qian, Dominic Rohner, Georgiy Syunyaev, Chris Udry, Paolo Zacchia, seminar participants at the Berlin Applied Micro Seminar, Boston University, Cardiff, Chicago Booth, Georgetown, EIEF, ENS Lyon, IEB, McGill, Michigan, MIT Sloan, Northwestern, Nova SBE, Stanford GSB, Strathclyde, and conference participants at UNCE~2019, EACES-HSE Political Economy Workshop, EWMES~2020, NEUDC~2020, ENS Lyon-CEPR ``Firm Behavior in Hostile Environments'' workshop, ArmEA~2021, SIOE~2021, IGIER-Bocconi Conference on Political Economy of Power Relations, AEA~2022, STEG Conference~2022, MPSA~2022, Naples Workshop on Networks and Development, CEPR-Misum-SITE Development Conference, and NBER Economics of National Security for useful feedback. We are grateful to Ella Sargsyan and Martin Strobl for superb research assistance. We thank the following entities for financial support: Czech Science Foundation (GACR grant \#19-25383S), Harriman Institute at Columbia University, and  UCLA Anderson Center for Global Management.}}\\[0.1cm]}

\author{
%\begin{tabular}{c p{1.5cm}c}
\begin{tabular}{c p{1.5cm} c}
Vasily Korovkin\thanks{CERGE-EI, a joint workplace of Charles University in Prague and the Economics Institute of the Czech Academy of Sciences, Prague, Czech Republic (e-mail: \href{mailto:vasily.korovkin@cerge-ei.cz}{vasily.korovkin@cerge-ei.cz}).} && Alexey Makarin\thanks{MIT Sloan School of Management, EIEF, and CEPR (e-mail: \href{mailto:makarin@mit.edu}{makarin@mit.edu}).}
\end{tabular}
}

\date{\normalsize First version: November~2020 \\ This version: July~2022 
%\textcolor{Blue}{Preliminary: please do not circulate.}
}

\vspace{-3cm}
\maketitle

\thispagestyle{empty}%
\setcounter{page}{0}

\vspace{-1.2cm}

\begin{abstract}
\begin{spacing}{1.0}
    
\noindent How do severe shocks such as war alter the economy? We study how a country's production network is affected by a devastating but localized conflict. Using unique transaction-level data on Ukrainian railway shipments around the start of the 2014 Russia-Ukraine crisis, we uncover several novel indirect effects of conflict on firms. First, we document substantial propagation effects on interfirm trade---trade declines even between partners outside the conflict areas if one of them had traded with those areas before the conflict events. The magnitude of such second-degree effect of conflict is one-fifth of the first-degree effect. Ignoring this propagation would lead to an underestimate of the total impact of conflict on trade by about \underestimationPerc. Second, war induces sudden changes in the production-network structure that influence firm performance. Specifically, we find that firms that exogenously became more central---after the conflict practically cut off certain regions from the rest of Ukraine---received a relative boost to their revenues. Finally, in a production-network model, we separately estimate the effects of the exogenous firm removal and the subsequent endogenous network adjustment on firm revenue distribution. For a median firm, network adjustment compensates for 80\% of the network destruction a year after the conflict onset.% and for all of it, two to three years removed. 

\end{spacing}
\end{abstract}
\vspace{-0.6cm}

\begin{center}
\begin{minipage}{0.92\textwidth}
\begin{small}
\end{small}
\end{minipage}
\end{center}
%\vspace{1in}
\end{titlepage}
\end{spacing}

\section{Introduction}
\label{SEC:Introduction}

\begin{spacing}{1.55}

%The more I think about it, the less sure I am about this motivations in the first paragraph:
%2.66 B people is fine, but can we also calculate GDP? Or some econ related measure?

International and civil wars have a devastating impact on the economy. Considerable work has been done to estimate the economic effects of conflict, at both the macro- and, more recently, the micro-level.\footnote{See, e.g., \cite{alesina1996political}, \cite{abadie2003economic}, \cite{cerra2008growth} for macro-level studies; see \cite{guidolin2007diamonds}, \cite{amodio2018making}, \cite{fisman2020experience}, \cite{ksoll2021electoral}, and \cite{korovkin2019trading} as well as \cite{blattman2010civil} and the references therein, for micro-level analyses of the economic effects of conflict.} Still, we know surprisingly little about how economic consequences of localized conflicts extend outside of the conflict areas. It is a significant knowledge gap given that, as of 2016, at least 2.66 billion people were living in countries with armed conflict but far from the actual violent events \citep{bahgat2018}. Moreover, if the spillover effects are nonzero, then the traditional methods of estimating the impact of conflict by comparing regions with and without violence \citep[e.g., ][]{abadie2003economic} may lead to biased estimates. 

We focus on the spillover effects of conflict on an often-neglected group of victims---firms. There are three ways in which conflict, or any other big persistent catastrophic event, affects firms not in the immediate vicinity of the shock. First, even keeping the production-network structure constant, firms may be indirectly hurt through a breakdown of their supply chain or an indirect decline in demand. Second, the destruction of firms and links, especially if the shock is sizeable, may alter the production-network structure, which may in turn affect a firm's relative network position and, as a result, its performance. Third, after the shock is realized, firms may readjust, finding new buyers and suppliers to form a new production-network equilibrium. Using a uniquely suitable shock and administrative data on firm-to-firm transactions, this paper provides---for the first time in the conflict literature and, for some effects, in economics more generally---estimates of all of the above effects.

We study these issues in the context of the Donbas War and the annexation of Crimea. Compared to other events of this kind, the Ukrainian conflict features three phenomena relevant to our argument. First, the conflict was isolated to a few territories next to the Russia-Ukraine border, and the risk of violence outside of Donbas was extremely limited throughout the period of our study. Second, Crimea and (especially) Donbas were economically crucial regions, jointly responsible for 17.5\% of Ukraine's GDP in 2013, so removing parts of those regions constituted a huge, persistent shock to the Ukrainian economy. Third, the availability of data on the universe of interfirm railway transactions from 2013 through 2016 allows us to study the dynamics of the Ukrainian production network both before and after the beginning of the conflict.

We start with the evidence on war shock propagation. In a difference-in-differences framework, we estimate the impact of the start of the conflict on interfirm shipments between the conflict and nonconflict areas (\textit{first-degree effect}), as well as on interfirm exchange in which both partners were outside the conflict areas but one of them had a trading partner located in the conflict areas before the conflict (\textit{second-degree effect}). Our estimates of the first-degree effect suggest that having one of the partners located in the conflict area reduces one's probability of monthly trade with that partner by 11.4 percentage points, equivalent to 0.43 standard deviations or 77\% of the preconflict mean. Trade intensity, measured in total weight traded and the number of transactions, reduces by 0.37 to 0.42 standard deviations or by 78--79\% relative to the preconflict mean. We find that demand shocks left bigger effects than supply shocks, i.e., that interfirm shipments decreased more when a customer and not a supplier was located in the conflict area. 

Interestingly, the second-degree effect is also negative. Its magnitude is about one-fifth of the first-degree effect's magnitude. However, due to it reducing the control group contamination and due to the large number of second-degree connections, ignoring the second-degree effects in a na\"ive difference-in-differences specification leads to underestimating the total effect of conflict on interfirm trade by \underestimationPerc. This result is in line with the findings for natural-disaster shocks in developed countries and aggregate firm outcomes \citep{barrot2016input,boehm2019input,carvalho2021supply}. However, this finding is not \textit{ex ante} obvious as, theoretically, the sign of the second-degree effect could be either positive or negative depending on the complementarity-substitutability patterns in the economy.\footnote{For example, trade between firms \textit{S} (seller) and \textit{B} (buyer) outside of the conflict areas could have gone up if a shock removed competitors of firm \textit{S}. An empirical question is, thus, which effect dominates in equilibrium.} In line with this argument, we find that having partner with a customer in the conflict areas is twice as harmful for inter-firm trade as having a partner with a supplier in the conflict areas.

Next, we show that the war shock alters the structure of the production network which, in turn, affects firm performance. That is, we find that firms that exogeneously gained in production-network centrality due to the start of the conflict have experienced a persistent increase in sales and a temporary rise in profits. We show this in a difference-in-differences framework, using firm-level administrative data and the network structure inferred from the interfirm shipments. To combat an obvious reverse-causality issue---firms that grew larger could automatically have become more central---we compute an \textit{exogeneous} change in a firm's network position based on its position in the preconflict production network. Specifically, we take the 2013 production network, artificially remove firms located in Crimea and the separatist part of Donbas, and recalculate firms' centrality in this new predicted network. We then show that this exogeneously altered centrality is highly predictive of a firm's actual realized centrality in 2014. Using this measure, we show that firms which, for exogenous reasons (i.e., due to a sudden
start of the Russia-Ukraine conflict cutting certain regions off from the rest of the country), became more central in the new production network also gained in sales and profits. A one-standard-deviation increase in a firm's network centrality led to an 11.5\%--14.5\% relative increase in its sales, depending on the centrality measure.
%\footnote{We consider three measures of centrality: eigenvector centrality, betweenness centrality, and degree centrality. While magnitudes vary, qualitatively our results are identical independent of the measure.} 
We find no evidence of pretrends, supporting the validity of the difference-in-differences strategy. In terms of the mechanisms, we find evidence suggesting that part of the centrality effect on firm performance may be coming from a short-lived increase in charged prices and market power.

Finally, we study how quickly Ukraine's production network readjusted after this large, persistent shock. To study this, we build a general equilibrium model in the spirit of the production-network literature\footnote{See \cite{acemoglu2012network,magerman2016heterogeneous,acemoglu2017microeconomic, baqaee2019macroeconomic,carvalho2021supply}.} with the CES preferences and production function. This framework allows us to achieve three goals: (i) switch from relative micro-level results to aggregate macro-level estimates; (ii) compare the magnitude of an exogeneous network-destruction channel and an endogeneous network-adjustment channel in a set of counterfactual exercises (specifically, the model yields a firm-level Leontief equation that allows us to manipulate the production-network matrix and obtain counterfactual estimates for the distribution of firms' revenues); and (iii) control for outside demand shocks. The latter also allows us to account for economic activity outside of the railway production network. 
%Note, that in the reduced form, outside demand shocks are absorbed by time and firm fixed effects, \textcolor{blue}{and network adjustment is shut down by our empirical strategy with a predicted network change.} 

%Finally, we separate the contributions of exogeneous network destruction and endogeneous network adjustments after the war shock. For this... Three factors determine the overall change of firm sales after the conflict. The first determinant is an exogenous network shock---conflict removes firms in conflict regions from the production network. The second determinant is the outside demand shock---the reduction of demand from final consumers and other customers outside of the railway data. The final, third determinant is the network adjustment of the firms---firms can search for new suppliers or customers after the shock. 

We conduct two counterfactual exercises, both of which compare firm-revenue distributions varying only the production-network structure and fixing the outside demand at the preconflict level. The first exercise considers a truncated production-network matrix, which consists of the 2013 production network but without the firms located in the conflict areas. The second network we consider is the one we actually observe in 2014. Conceptually, the first matrix includes only the exogeneous link destruction due to the war, while the latter one allows for endogenous link formation that occurs afterward. Fixing the outside demand at the preconflict level, the revenue distribution under a counterfactual scenario with the first network gives us an estimate of what we call a \textit{network-destruction channel}. Similarly, the difference in revenue distributions under the two counterfactual scenarios gives us an estimate of the \textit{network-adjustment channel}.

Estimating these two counterfactuals on the Ukrainian data suggests that, on average, the economy managed to adjust to the shock relatively quickly but not in full. Specifically, while the network-destruction effect decreases median firm sales in nonconflict areas by 46.8\%, network adjustment alleviates 80\% of this drop. At the same time, consistent with our reduced-form centrality results, bigger firms (75th percentile in revenue) managed to fully readjust and even gain from the total change in the production-network structure, increasing their sales in the adjustment counterfactual scenario by 14\%. In contrast, at the left tail of the distribution (25th percentile), network adjustment compensates for only 33\% of network destruction. Overall, these results provide a novel quantification of the role of network adjustment after a persistent shock and indicate that firms in nonconflict areas readjusted relatively rapidly and that such shocks may have increased interfirm inequality.

Our article builds on the literature on the economic effects of conflict.\footnote{We stand on the shoulders of a vast literature on conflict in economics and political science. Seminal contributions include but are not limited to \cite{esteban1999conflict, fearon2003ethnicity,montalvo2005ethnic,nunn2014us} and \cite{dell2017nation}. See \cite{blattman2010civil} and \cite{rohner2020elusive} for the in-depth overviews of this literature.} %The closest paper to ours, \cite{amodio2017making}, shows how conflict changes the input choices of Palestinian firms and creates misallocations. 
We contribute, first, by documenting that the conflict shock has negative economic spillovers on localities and firms outside the conflict areas by propagating through the production network. This finding casts doubt on the validity of the difference-in-differences and synthetic control comparisons of violent and nonviolent areas for gauging the economic cost of conflict (e.g., \citealp{abadie2003economic}).
%\footnote{For the most recent review of the literature on the economic costs of war, see \cite{rohner2020elusive}.} 
Second, we show that sufficiently persistent violence may alter the structure of the production network benefiting some firms in the rest of the country, but that endogeneous network adjustment helps the nonconflict areas recover relatively quickly. These novel findings add to the strand of research that studies effects of conflict on firms at the micro level \citep{guidolin2007diamonds, amodio2018making,utar2018firms,ksoll2021electoral,del2021firms}. In contrast, the economic spillovers of conflict on firms are understudied, with the exception of \cite{hjort2014ethnic} and \cite{korovkin2019trading}, who show how conflict-induced intergroup tensions affect, respectively, firm productivity and interfirm trade outside the conflict areas.\footnote{See also \cite{amodio2021security} who document the indirect effects of conflict on firms due to security-motivated trade restrictions. Finally, a related strand of the literature concerns the relationship between conflict and trade \citep{martin2008civil,martin2008make,guiso2009cultural,rohner2013war}.} Contemporaneous work by \cite{couttenier2022economic} documents the spillovers of Maoist conflict in India on firms through a gravity-predicted network of trade connections.
%We contribute to the existing studies by analyzing the impact of conflict on domestic, as opposed to international, trade using micro-level data on interfirm shipments.}
%\textcolor{blue}{Arguably, network distance can be more important than geographic distance in the context of conflicts.}
%Moreover, as a result, we can build a macro estimate from the micro-level shipment data.

%We add to the literature on networks, trade, and firm performance. Due to the uniqueness of our research context and data, we are the first to use an exogenous change in production-network structure to estimate the impact of firm's network position on firm performance. For a comprehensive review of the literature on networks and trade, see \cite{bernard2018}. The closest paper to ours is \cite{bernard2019production} which uses an opening of a high speed railway line in Japan on firm performance through finding additional customers and suppliers. In contrast, we study how an abrupt change in the production network of the whole economy may actually benefit some firms by making them more central relative to others.\footnote{Other related papers include \cite{wu2014supply,herskovic2018networks,zacchiaknowledge}} \textcolor{blue}{What is equally important, is that the shock, coming from the conflict is more exogenous than railways construction.}

We also contribute to the literature on production networks. Due to the unique features of our data and context, we are able to provide the rare empirical analysis of the endogenous production network readjustment after a large negative shock. Furthermore, the sudden and persistent nature of the shock allows us to study the impact of the corresponding exogeneous changes in the production-network structure on firm performance. The literature on production networks generally stresses the importance of input-output linkages for propagation and amplification of shocks \citep[e.g.,][]{acemoglu2012network,bigio2016financial,acemoglu2017microeconomic,baqaee2018cascading,baqaee2019macroeconomic,baqaee2020productivity,ferrari2019global,carvalho2021supply}. \cite{acemoglu2020endogenous} present a theoretical model of endogenous production network formation. On the empirical side, \cite{barrot2016input}, \cite{boehm2019input}, and \cite{carvalho2021supply} study the propagation of shocks within the networks of suppliers and customers on aggregate firm outcomes in the context of natural disasters in the United States and Japan. We complement these studies by documenting propagation effects of a large persistent shock---in this case, war---using granular data on trade intensity. \cite{liu2020dynamical} study the welfare effects of temporary shocks in a dynamic setting with fixed production-network structure, applying their results to bombing-target selection during WWII. Our study is complementary: we provide the first analysis of a reverse causal link, i.e., how conflict affects production networks.

%We are the first to use a large and exogenous shock---a war---typical for both developing countries and throughout the history of humankind---and apply it to a production network. More importantly, our setting allows us to manipulate the input-output network and provide relevant counterfactuals for both this large exogenous shock and firms' reaction to the war onset.

%Papers in international and interregional trade emphasize that input-output linkages can make business cycles in different countries correlated with each other and can propagate changes in productivity within the U.S. \citep{johnson2014trade, di2018micro,caliendo2017impact}. Closer to our setting, three works by \cite{barrot2016input}, \cite{boehm2015input}, and \cite{carvalho2021supply} who use natural disasters in the USA and Japan to measure the propagation of shocks within the networks of suppliers and customers, and from Japan to the American affiliates of the Japanese firms. Importantly, our paper is the first to our knowledge to apply these insights to a conflict setting and outside of the developed world.

The rest of this paper is organized as follows. \Cref{sec:background} describes the Ukrainian economy before the 2014 conflict, as well as the details of the conflict's onset. 
%\Cref{sec:model} builds a stylized model of production networks. 
%\Cref{sec:empstr} explains how we transition from a theoretical model to our main empirical exercises.
\Cref{sec:data} discusses the data and provides descriptive statistics. \Cref{sec:results_propagation} presents the propagation results. \Cref{sec:results_centrality} explores the impact of an exogeneous change in production-network structure on firm performance. \Cref{sec:concframework} discusses the conceptual framework and compares the losses from exogenous network shock to the losses if the network adjustment channel is allowed to operate.
%\Cref{sec:robustness} presents robustness checks. 
\Cref{sec:conclusion} concludes.

%\newpage
\section{Background}
\label{sec:background}

\subsection{Annexation of Crimea, and the Donbas War (2014--2022)}

%In November 2013, Victor Yanukovych, the former president of Ukraine and a loyal ally of Russia, walked back his promise to enter a political and economic association with the European Union. This decision led to a massive wave of protests across the country. After several deadly clashes between protesters and the police, Victor Yanukovych fled to Russia on February~22,~2014, and, at that point, the revolution had succeeded. 

Immediately after the Ukrainian revolution of February~2014, the Russian government decided to occupy Crimea and started promoting separatist movements in the Donetsk and Luhansk provinces (i.e., the Donbas region).\footnote{The decision on Crimea was made secretly by Vladimir Putin and a handful of senior security advisors. It took everyone else by surprise \citep{treisman2018new}.} The annexation was complete by early March~2014; it occurred without direct military conflict. Later, pro-Russian protests ensued in Donbas. Parts of these areas proclaimed independence from Ukraine, forming the Donetsk People's Republic (DPR) on April~7, 2014, and the Luhansk People's Republic (LPR) on April~27, 2014. In response, the new acting Ukrainian president launched an ``Anti-Terrorist Operation'' operation to suppress these separatist movements. Russia supported the DPR and LPR and, among other things, provided them with military power. A long-lasting conflict ensued, leading to more than 13,000 casualties, 30,000 wounded, and the displacement of hundreds of thousands of people.\footnote{E.g., see \url{https://neweasterneurope.eu/2019/09/24/the-cost-of-five-years-of-war-in-donbas/}} The conflict has been in a rather ``frozen'' state ever since the election of President Zelensky. That abruptly changed on February 24, 2022 when Russia launched a full-scale invasion of Ukraine.
%Using nighttime luminosity data, researchers have documented that the separatist rule led to a 40--70\% economic decline in the Donbas area \citep{coupe2016lights,kochnev2019dying}.

\begin{figure}[!t]
\begin{center}
\begin{minipage}{\textwidth}
    \begin{center}
        \includegraphics[width=\textwidth]{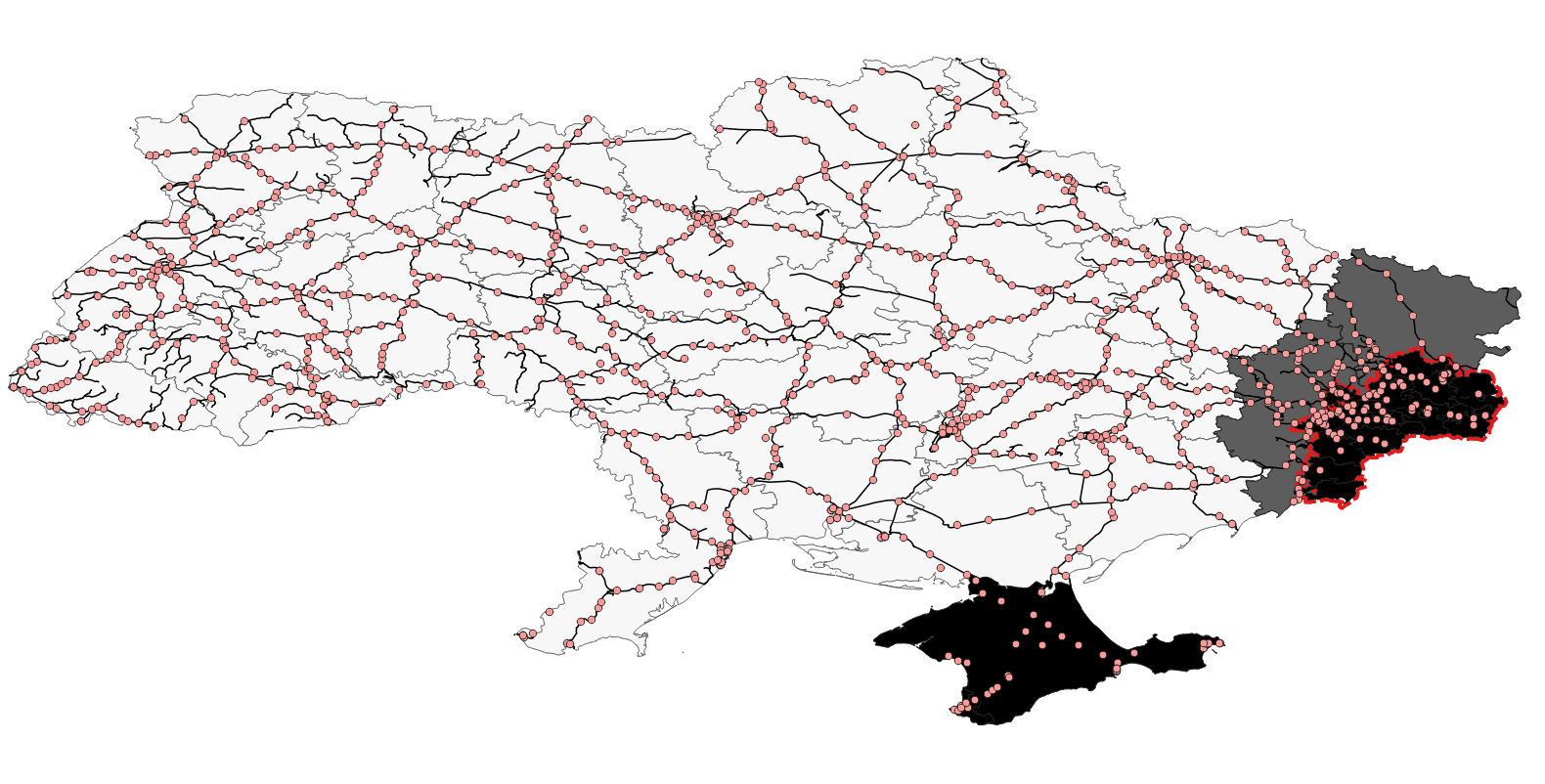}\\
    \end{center}
    \vspace{-0.2cm}
\caption{Conflict Areas (2014--2022) and Railroads in Ukraine}\label{FIG: Map 1} %
\vspace{-0.2cm}
\renewcommand{\baselinestretch}{0.7}
\footnotesize{\textit{Notes}: The map highlights the areas directly affected by the 2014 Russia-Ukraine conflict and displays the geographic location of the railroads and the railway stations. The Crimean Peninsula, in black at the bottom of the map, was occupied by Russia in early 2014. The Donetsk People's Republic (DPR) and Luhansk People's Republic (LPR) territories are in black with a red rim to the right. The rest of the Donbas region, in light gray, consists of the Donetsk and Luhansk provinces. Black lines indicate the Ukrainian railroads. Red dots represent the railway stations in our railway-shipments data.}
\end{minipage}
\end{center}
\vspace*{-0.2cm}
\end{figure}

\Cref{FIG: Map 1} shows the areas directly affected by the 2014 Russia-Ukraine conflict. These include Crimea (in black at the bottom) and the two quasi-independent states of the Donetsk and Luhansk People's Republics (in black with a red rim, on the right side of the map). While the conflict was intense in certain DPR and LPR territories, especially at their respective borders, the rest of the country was not exposed to violence directly.

%Figure 1

\subsection{Economic Activity in Donbas and Crimea}

Before the conflict, Donbas and Crimea were critically important for Ukraine's economy. Together, they accounted for about 17.5\% of Ukraine's 2013 GDP. The Donbas region has always been prominent for its extractive industries, especially coal, metallurgy, and manufacturing. Being the most populous province in Ukraine, with 4.4 million people, or 10\% of Ukraine's population, Donetsk oblast (province) has been responsible for more than 20\% of all Ukrainian manufacturing and 20\% of all Ukraine's exports as of 2013. 

Though less economically important than Donetsk, Luhansk oblast (province) has also been essential for Ukraine; it was the sixth-most-populous Ukrainian province, with 2.16 million people producing 6\% of Ukraine's exports. In contrast to Donbas, Crimea is particularly well-known for its agricultural products and tourism. However, it was also a vital part of Ukraine's economy before the conflict, hosting 2.2 million people and being the center of several industries, such as shipbuilding.

The consequences of conflict for these regions were devastating. Crimea became almost entirely cut off from the Ukrainian transportation network, leading to a sudden disruption of supply-chain links. The DPR and LPR were overtaken by violence, bombing, destruction of infrastructure and physical capital, and the loss of labor force. In the course of a year, manufacturing production fell by 60\% in Donetsk oblast and by 80\% in Luhansk oblast. Due to the complexity of the conflict treatment, we will not separately estimate the impacts of each of its components and instead focus on the aggregate consequences of the conflict shock.
%This paper provides the first rigorous estimates for how this shock affected the rest of Ukraine's economy.

\subsection{Ukrainian Railroad System}

Railway transportation is critical for Ukraine's economy. Ukraine has the 13th-most-extended railroad network and is the world's seventh-largest railway freight transporter. Railroads are the main way of transporting products in Ukraine: according to UkrStat, as of 2018, railroads were responsible for 80\% of ton-km of all freight transport.\footnote{\url{http://www.ukrstat.gov.ua/operativ/operativ2018/tr/vtk/xls/vtk_2018_e.xlsx}.} Meanwhile, other modes of transportation are not particularly well maintained. According to the WEF Global Competitiveness Report, the Ukrainian railroad infrastructure is among the best in the world (25th in 2013--2014).\footnote{\url{https://www.weforum.org/reports/global-competitiveness-report-2013-2014}.} In contrast, regular roads and airway transportation are poorly ranked relative to those in other countries (144th and 105th in the world in 2013--2014, respectively). 

%Ukrainian railroads are 6th lengthiest in Europe and 12th lengthiest in the world. Ukraine hosts 1,614 railway stations, of which 1,200 appear in our data. In total, Ukrainian railway company has about 132,500 railway wagons and employs 385,000 people, which makes it the largest company in Ukraine. 

%In general, Ukrainian railway system and wagons are in a poor technical state.

%\section{Model}
%\label{sec:model}

%\section{Empirical Strategy}
%\label{sec:empstr}

\section{Data}
\label{sec:data}

In our analysis, we rely on two main datasets. First, we employ novel data on the universe of railroad shipments within Ukraine from 2013 through 2016.\footnote{In focusing on railroad transportation and trade, we relate to \cite{donaldson2016railroads} and \cite{donaldson2018railroads} who study the economic impact of railroads using historical data on the agricultural sector in the United States and India, respectively.} These granular data allow us to study the evolution of a production network before and after the start of a large-scale conflict.  The dataset contains more than 41 million transactions between more than 7,000 firms. It includes shipment dates, weights (in kg), freight charges, product codes, and station codes filled out my railway clerks. Importantly, the dataset contains unique IDs of the Ukrainian firms-senders and firms-receivers, which enables us to merge the dataset with other firm-level data. 

Since Ukrainian railways operate through a state monopoly, no inter-firm railway shipments pass without entering these records. However, to further account for possible mechanical disruptions in data reporting due to the conflict, we exclude the railway shipments conducted fully within the separatist-controlled territories from all of our analysis.

Throughout the paper, we also discard railway shipments passing Ukraine in transit. In contrast, in order to consider all possible alternative buyers and suppliers, we include the Ukrainian firms' international railway shipments. This choice does not qualitatively affect our estimates as they remain robust to excluding international trade and focusing only on domestic shipments.   

Second, we use accounting data from \cite{sparkinterfax}. SPARK-Interfax is similar to the ORBIS/AMADEUS database but has a strong focus on the former Soviet republics, including Russia, Ukraine, Kazakhstan, Belarus, Kyrgyzstan, Uzbekistan, and Moldova. The dataset contains information on firm sales, profits, total costs, capital, and other variables for more than 370,000 Ukrainian firms from 2010 through 2017. Similar to ORBIS/AMADEUS, SPARK-Interfax is based on official government statistics, provision of which is mandatory for all Ukrainian firms except individual entrepreneurs.\footnote{As noted in \cite{kalemli2015construct}, Ukrainian filing requirements are among the most demanding in the world. We are unaware of any estimates of the SPARK-Interfax or ORBIS/AMADEUS coverage for Ukraine, but in Romania, a neighboring country with similar filing requirements, ORBIS/AMADEUS was found to cover 92\% of gross output and 93\% of total employment in the manufacturing sector \citep{kalemli2015construct}.}

\Cref{TABLE: Summary Statistics} displays the summary statistics for both datasets. On average, a firm sent a shipment to another firm every 13 months. The average weight of the monthly shipments was 257 tons, equivalent to seven railway wagons of coal. In 10.5\% of firm pairs, one of them shipped or received either in Donbas or Crimea, and in 62.5\% of firm pairs, one of them had a trading partner located either in Donbas or Crimea. %
%\footnote{The fact that this production network saturates so quickly prevents us from looking at the third- and fourth-degree propagation of the war shock.} 
Finally, the median total revenue and profits of the firms matched with the SPARK-Interfax dataset over the 2010--2017 period was 26.9 million and 260 thousand Ukrainian hryvnia (around US\$1,000,000 and US\$10,000 as of November 2021), respectively.

\Cref{FIG: Map 1} depicts the Ukranian railway network, as well as the 1,200 railway stations in our dataset. The stations cover the entire territory of Ukraine, indirectly confirming the universal nature of our data. As one can see, the railway network is especially dense in the Donbas region. This pattern is consistent with the Donbas' heavy reliance on railway transportation, given its focus on coal and mineral extraction, metallurgy, and other heavy industry.

%\section{War and Trade within Ukraine} 
%\label{sec:results}

%We start by documenting a large and significant decline in trade between provinces if a province is affected by combat. We then proceed to documenting propagation of the war shock through the production network. To make a causal claim, we use a difference-in-differences specification. We then decompose the effects by sending and receiving firms to study the differences in upstream and downstream propagation.

%In the second step, we focus on the changes in network centrality, and how those changes affect firm performance. We build an exogenous component of the change in centrality and use it instead of the observed change in centrality from 2013 to 2014. %We use this ``predicted'' change in centrality in a difference-in-differences specification to measure the effect on firm yearly sales and profits.

\section{War and Propagation Through Trade Linkages}
\label{sec:results_propagation}

In this section, we establish our first main result: the conflict shock not only affects trade between firms directly exposed to violence but also propagates---through the production-network linkages---to trade between firms outside the conflict areas. This result is key for the literature on the economic consequences of wars and violence: it suggests that the standard techniques for comparing conflict and nonconflict areas (e.g., \citealp{abadie2003economic}) may severely underestimate the total economic impact of these events.

To document this result empirically, we estimate a difference-in-differences specification with interfirm trade intensity as an outcome variable\footnote{This specification is in the spirit of the production-network literature (see, e.g., \citealp{carvalho2021supply}). However, due to data limitations, previous studies have focused on aggregate outcomes \citep{carvalho2019production}, such as yearly sales, and not the disaggregated trade flows, as in our case. Still, we present the propagation estimates on firm sales in Figure~\ref{FIG: Sales, Network Centrality and Conflict} (discussed further in Section~\ref{sec:results_centrality}).} and with measures of direct and indirect connections to the conflict areas on the right-hand side of the equation. 

We leverage detailed transaction-level data to precisely identify trade links that were directly affected by the annexation of Crimea and the Donbas War. Specifically, we consider the geographic origin of the trade-link nodes, and we study trade between firm-rayon pairs.\footnote{Ukrainian rayon (district) is a subnational administrative unit below the level of oblast (province). At the time of the study, Ukraine was divided into 25 oblasts and 490 rayons.} We consider a particular trade link ``first-degree-treated'' (i.e., affected directly) if one of the firms was shipping or receiving products directly from or in a conflict-affected rayon. Similarly, we consider a certain trade direction between two firms as ``second-degree-treated'' (i.e., affected indirectly) if one of the firms traded with the conflict-affected rayons before the start of the conflict, and did so from the same location that they used to ship products within the given trade link. One can think of this approach as performing an establishment-level analysis.\footnote{In Table~\ref{TABLE: Propagation Firm Level Aggregation}, we show that our results are robust to defining the second-degree treatment effect at the firm level instead of at the establishment level, thus accounting for within-firm shock propagation or absorption.} 
%An intuitive interpretation of this approach is that we study trade between establishments and do not consider within-firm spillovers. 

With these caveats in mind, we estimate the following specification:
\vspace*{-0.25cm}
\begin{equation} \label{eq: propagation}
Y_{ijt} = \alpha_{ij} + \kappa_t + \beta^{(1)} \times \textnormal{Conflict}_{ij} \times \textnormal{Post}_t + \beta^{(2)} \times \textnormal{PartnerConflict}_{ij} \times \textnormal{Post}_t  + \eta_{ijt}
\vspace*{-0.25cm}
\end{equation}
where $Y_{ijt}$ represents trade intensity from establishment (firm-rayon) $i$ to establishment $j$ in a given year-month $t$,\footnote{The specification allows for trade in the opposite direction---e.g., $Y_{jit}$ would instead indicate trade from $j$ to $i$.} measured either as probability of trade, log of the number of shipments, or log of the total weight of transactions within a trade link; $\textnormal{Post}_t$ is the post-Crimea indicator; $\textnormal{Conflict}_{ij}$ is an indicator for whether the trade link is directly affected by conflict, equal to one if either establishment $i$ or $j$ is in DPR, LPR or Crimea (and zero otherwise); $\textnormal{PartnerConflict}_{ij}$ is an indicator for whether the trade link is affected by conflict one-degree-removed and is equal to one if either $i$ or $j$ had a preconflict partner (other than $j$ or $i$) that was located in DPR, LPR or Crimea and equal to zero otherwise or if either establishment was itself located in DPR, LPR or Crimea;
%\footnote{There are practically no trade links in which one of the partners was in the conflict area but neither had traded with the conflict areas before the conflict. We exclude trade within conflict areas from our analysis.} 
$\alpha_{ij}$ and $\kappa_t$ are the establishment-pair(-direction) and year-month fixed effects, respectively. Under an assumption that trade intensity in links with and without first- and second-degree ties to conflict areas evolves along parallel trends, the coefficients $\beta^{(1)}$ and $\beta^{(2)}$ identify the first- and second-degree effects of conflict on trade.\footnote{\cite{borusyak2020non} argue that in similar settings the treatment effects estimators $\hat\beta^{(1)}, \hat\beta^{(2)}$ could be biased because the ``periphery'' firms are less likely to be affected by any network-based shock. In Table~\ref{TABLE: Propagation Borusyak and Hull Adjustment}, we present robustness to one of the remedies suggested by the authors and control for number of partners each firm had preconflict interacted with the postconflict indicator. The estimates remain largely unchanged.}\footnote{We acknowledge there could also exist the third- and higher-degree effects of conflict. We do not consider further propagation for three reasons. First, our network saturates quickly. There are practically no trade links with no first, second, or third-degree connections to the conflict areas, which implies lack of a proper control group for the higher-degree effects. Second, to reject the null hypothesis of no spillovers, it suffices to find substantial second-degree effects. Finally, if the third or higher-degree effects are also negative, then the $\beta^{(1)}$ and $\beta^{(2)}$ estimates represent a lower bound of the first- and second-degree effects.}

We report the estimated coefficients in Table \ref{TABLE: Reduced-Form Estimates of the Reduction in Trade}, starting with the first-degree effects only and then adding the second-degree effects. Across all specifications, the first- and second-degree coefficients are negative, statistically significant at the 0.1\% level, and of substantial magnitude. 

In our preferred specifications, in column (4)--(6), the comparison group consists of the trade links in which no trading partner was in the conflict area and also none of the trading partners had traded with the conflict areas before the start of the conflict. Adding an indirect connection with the conflict---that is, a link through a buyer or a supplier---dampens the monthly probability of trade between two establishments by 2.5 percentage points, a sizeable decline, equivalent to 0.09 standard deviations. If, instead, a trade link was directly connected with the conflict, the monthly probability of trade falls by 13.1 percentage points or 0.49 standard deviations. The results for the log number of shipments and log-weight are qualitatively similar, suggesting that trade volume declines in similar proportions to trade frequency. 

%Table 1
\begin{table}[!t]\centering
\caption{Propagation of the Conflict Shock Through Interfirm Trade Linkages}\label{TABLE: Reduced-Form Estimates of the Reduction in Trade}
\vspace{-0.1cm}
\resizebox{\textwidth}{!}{%
\begin{tabular}{lcccccc}
\hline
			&       (1)       &       (2)        &       (3)        &       (4)        &       (5)        &       (6)  \\
           &       Any        &       Log       &       Log Total        &       Any        &       Log       &       Log Total     \\  
           &       Shipment        &      Number of      &       Weight        &      Shipment       &     Number  of       &       Weight     \\  
            &              &      Shipments        &       Shipped        &            &    Shipments         &       Shipped     \\  
           \hline
           \\[-0.25cm]
           
\input{Tables/Table_1_Firm_Rayon_Month}

\hline \hline 
\end{tabular}}
\\
\begin{minipage}{\textwidth}
\vspace{0.1cm}
\renewcommand{\baselinestretch}{0.7}
\footnotesize{\textit{Notes:} This table presents the estimates of equation~\eqref{eq: propagation} studying whether interfirm trade declined after the start of the Russia-Ukraine conflict depending on whether one of the partners in an exchange was located in or had traded with the conflict area before the conflict. The outcomes are: an indicator for any shipment from establishment (firm-rayon) $i$ to establishment $j$ at year-month $t$, the logarithm of the total number of shipments from establishment $i$ to establishment $j$ at year-month $t$, and the total weight of these shipments. To avoid missing values, the logarithms add 1 to the value of the argument. Standard errors in parentheses are clustered at the province-pair level. * p$<$0.05, ** p$<$0.01, *** p$<$0.001.\\
}
\end{minipage}
\end{table}

There are two ways in which na\"{i}ve specifications in columns (1)--(3) underestimate the total impact of conflict on trade. First, the negative second-degree effect contaminates the comparison group leading to a downward bias. Indeed, once we allow for their presence, the first-degree-effect coefficients grow in magnitude---e.g., from $-$0.114 in column~(1) to $-$0.131 in column~(4) for the monthly probability of trade. Second, the presence of negative second-degree effects, even if small, significantly alters the aggregate calculations of the impact of conflict. This is because the second-degree links with the conflict areas are much more numerous than the first-degree links. In sum, when we aggregate the coefficients up by the number of first- and second-degree connections to the conflict areas, we find that ignoring the second-degree effects leads to underestimation of the total impact of conflict on trade by 65\% to 67\%, depending on the measure of trade intensity.

\begin{figure}[!t]
\begin{center}
\begin{minipage}{0.98\textwidth}
    \begin{center}
        \includegraphics[width=\textwidth]{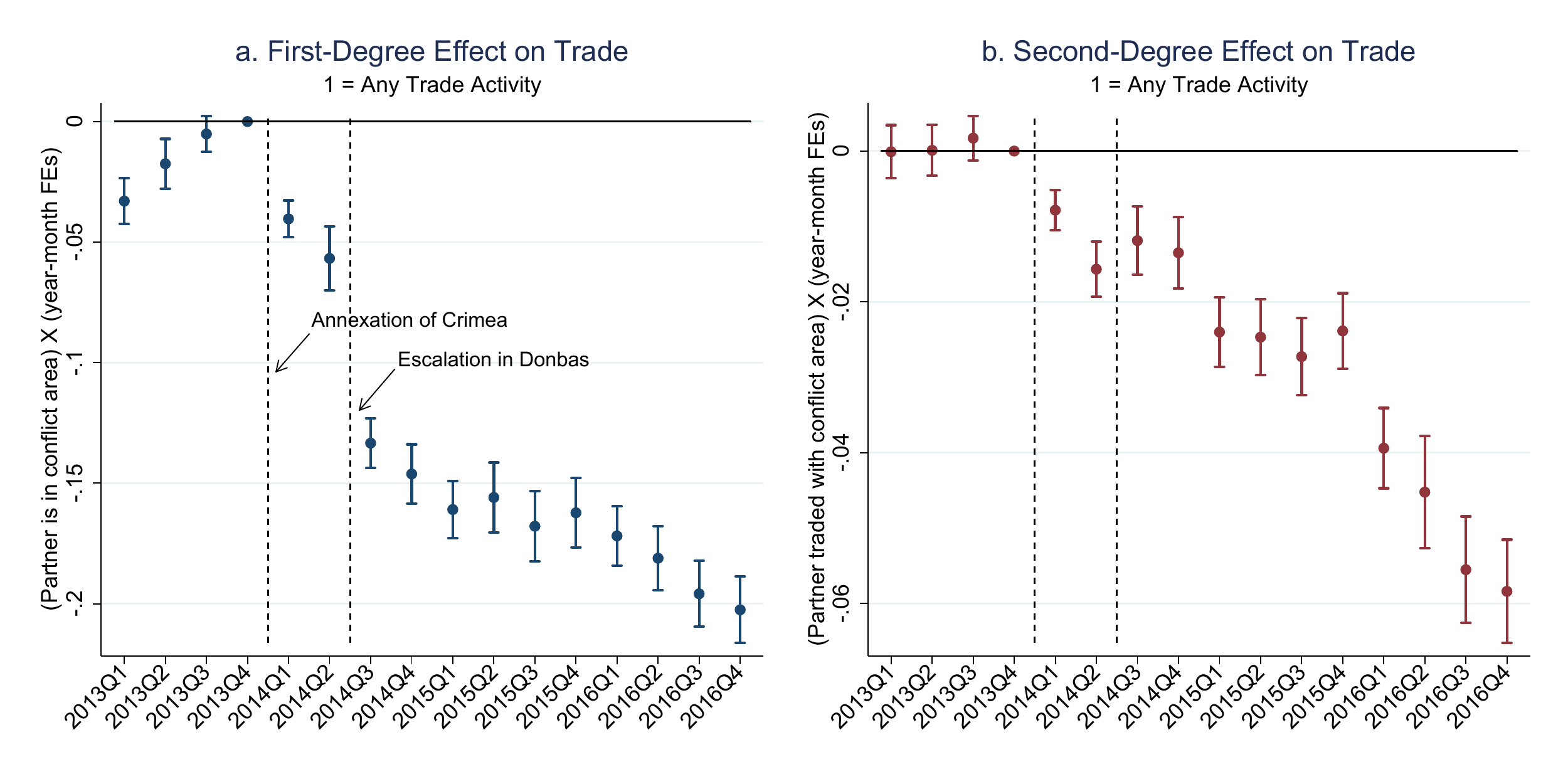}
    \end{center}
\caption{Propagation Effects of Conflict on Trade}\label{FIG: EventStudy_FirstPart} %
\vspace{-0.2cm}
\renewcommand{\baselinestretch}{0.7}
\footnotesize{\textit{Notes}: The figure displays the results of estimating a quarter-by-quarter version of equation \eqref{eq: propagation} where first- and second-degree are both interacted with the quarterly indicators. Panel A displays the results for the first-degree (direct) effects of conflict on trade links in which one of the partners was located in the conflict area. Panel B displays the results for the second-degree (indirect) effects of conflict on trade links in which one of the partners had traded with a firm in the conflict areas before the start of the conflict. Bars represent 95\% confidence intervals. Standard errors are clustered at the province-pair level. $N=11,001,648$.}
\end{minipage}
\vspace*{-0.5cm}
\end{center}
\end{figure}

Figure~\ref{FIG: EventStudy_FirstPart} illustrates the effects' dynamics. Specifically, it presents the estimates of the equation~\eqref{eq: propagation} but with $\textnormal{Conflict}_{ij}$ and $\textnormal{PartnerConflict}_{ij}$ interacted with the quarterly indicators instead of the post-Crimea indicator. Panel~A displays the dynamics of the first-degree effects, and Panel~B displays the second-degree effects. The last quarter of 2013 serves as a baseline. Despite some pretrends, Panel~A clearly shows that the annexation of Crimea, and especially the escalation of violence in Donbas, has led to a sharp decline in the first-degree trade with the conflict areas.
%\footnote{Table~\ref{TABLE: Propagation Crimea Donbas} presents the results separately for Crimea and Donbas, both in terms of firm connections and timing. }
%\footnote{We speculate that the pretrends for the first-degree effects have to do with the seasonal patterns in trade between Crimea, Donbas, and the rest of Ukraine.} 
The second-degree effects displayed in Panel~B exhibit no pretrends before the conflict, lending strong support for the parallel-trends assumption in this case. The coefficients' pattern suggests that conflict also has sharp and deep negative indirect consequences appearing immediately with the annexation of Crimea and slowly increasing in magnitude over time.
%\footnote{We speculate that the lack of an immediate decline after the escalation in Donbas might be explained by storage and gradual deterioration of supply chains.}

%Figure 4

Next, we compare upstream and downstream propagation by splitting the data further into having a buyer or a supplier directly or indirectly connected to the conflict areas.\footnote{In particular, we estimate the following specification: %
\begin{align} \label{eq: propagation separated}
\begin{split}
Y_{ijt} = &\alpha_{ij} + \kappa_t + \beta_b^{(1)} \times \textnormal{BuyerConflict}_{j} \times \textnormal{Post}_t + \beta_s^{(1)} \times \textnormal{SupplierConflict}_{i} \times \textnormal{Post}_{t} + \\
&\beta_b^{(2)} \times \textnormal{PartnerBuyerConflict}_{ij} \times \textnormal{Post}_t +
\beta_s^{(2)} \times \textnormal{PartnerSupplierConflict}_{ij} \times \textnormal{Post}_t  + \eta_{ijt}
\end{split}
\end{align}
where $\textnormal{BuyerConflict}_{j}$ is an indicator for whether buyer $j$ is in DPR, LPR or Crimea; $\textnormal{SupplierConflict}_{i}$ is an indicator for whether supplier $i$ is in DPR, LPR or Crimea;  $\textnormal{PartnerBuyerConflict}_{ij}$ is an indicator for whether, before the start of the conflict, either partner $i$ or $j$ had a buyer located in DPR, LPR or Crimea; and $\textnormal{PartnerSupplierConflict}_{ij}$ indicates whether, before the start of the conflict, either partner $i$ or $j$ had a supplier located in DPR, LPR or Crimea.}
Table~\ref{TABLE: Propagation Downstream and Upstream} presents the estimates. In contrast to prior evidence in other contexts (e.g., \citealp{carvalho2019production}), the effects indeed differ by the direction of trade. Namely, the first-degree effects are larger if, of the two firms, a buyer was located in the conflict areas (columns~1 through~3). We interpret this disparity as the conflict-induced demand shocks being more severe compared to the supply shocks. For the second-degree effects, the pattern is similar: conflict-induced demand shocks propagate to second-degree links at a higher rate than the conflict-induced supply shocks (columns~4 through~6). One interpretation could be that supply shocks are less severe because of a possible substitution to competing inputs, while the search for new customers in the case of demand shocks is more difficult. Nevertheless, both demand and supply shocks still have sizeable negative second-degree effects, leading to declines in trade frequency in the indirectly connected links of 0.04 and 0.09 standard deviations, respectively.

One may wonder whether the above results are specific to certain industries more prevalent in Donbas or that more heavily rely on the railways as a mode of transportation. To discern the validity of this concern, we present the heterogeneity estimates by firms' industry. To obtain these estimates, we merge the transaction-level data with each firm's Standard Industrial Classification (SIC) codes and take the main SIC code listed for each firm. Then, following the official SIC manual \citep{osha2014sic}, we group the firms into coarse groups of industries, such as mining and manufacturing. Figure \ref{FIG: Propagation Heterogeneity by Industry} presents the heterogeneity estimates. The coefficients are consistent with the intuition that indirect effects should matter most for downstream sectors, such as retail, and matter least for upstream sectors, such as mining. Overall, we note that the degree of heterogeneity is relatively small and, thus, our baseline results are more likely to extend to settings with a different composition of industries and industries with varying reliance on railroad transportation.

Finally, one may be concerned that the results are due to some omitted variable shifted by the conflict and not the production-network propagation. For instance, one may worry that firms with the second-degree connections to the conflict areas could locate geographically closer to the violent areas and, thus, may be more likely to be hit by the concurrent conflict-related demand or supply shocks. Alternatively, second-degree connected firms may locate in provinces of Ukraine that experience relatively more negative indirect effects due to the conflict---e.g., they could host fewer refugees, who present positive labor supply and demand shocks to the local economy. Furthermore, ethnic and historical differences between Donbas, Crimea, and other Ukrainian provinces could have played a role in the post-conflict breakdown of trade \citep{markevich2021political,korovkin2019trading}. We address these concerns with two additional robustness checks. First, we flexibly control for the firm's distance to the conflict areas interacted with the post-Crimea indicator (Table~\ref{TABLE: Propagation Controlling For Distance}). Second, we control for the province fixed effects interacted with the post-Crimea indicator (Table~\ref{TABLE: Propagation Controlling For Province FE}). The results remain qualitatively unchanged. 

To summarize, using unique data on within-country trade transactions, we show that the start of a large-scale conflict had a negative impact not only on trade directly associated with the conflict areas but also on trade links connected with them only indirectly. The shock propagates both upstream and downstream, although the demand shock produce larger negative effects. Ignoring the propagation effects leads to severe underestimation of the total economic impact of conflict.

\section{War and Firm-Level Effects of Production-Network Structure Change}
\label{sec:results_centrality}

%\subsection{Induced Changes in Network Centrality and Firm Performance}

%In this section, we aim to show that war, as well as potentially other large-scale persistent shocks, induces a change in the production-network structure, which itself has an effect on firm performance.

In this section, we aim to show that a large-scale conflict does not only propagate through the production network but also, if the territorial breakaway is large and persistent---which was the case during the 2014 Russia-Ukraine conflict,---it can change the structure of the production network itself. In turn, this sudden change may have implications for firms in non-conflict areas due to an abrupt shift in their network position.

We illustrate this point empirically by estimating the impact of an exogenous conflict-induced change in firm centrality on several measures of yearly firm performance. A na\"{i}ve approach would regress the change in firm outcomes on the change in firm centrality between 2013 (before the conflict) and  2014 (after the start of the conflict). However, this regression may suffer from reverse causality---firms with a steeper rise in performance may automatically become more central since they would likely exhibit an increase in the number of transactions and trade with more firms. To preempt this and other similar concerns, we rely on an exogeneous component of the conflict-induced change in firm centrality based on the 2013 production network only. Specifically, we take the 2013 production network, drop the firms in the conflict areas from the network, and recalculate measures of centrality for the remaining firms. Figure~\ref{FIG: First Stage} illustrates that such predicted firm centrality is highly correlated with the actual firm centrality in 2014. We then take the difference between the predicted measure and the actual firm centrality in 2013. We then estimate the following difference-in-differences equation on levels of firm performance using the exogeneous war-induced change in centrality as the main independent variable:%
\vspace*{-0.25cm}
\begin{equation} \label{eq: centrality}
		Y_{it} = \alpha_i + \delta_t + \beta_{t} \times \Delta \widehat{\textnormal{Centrality}}_{i,2013} + \varepsilon_{it}
\vspace*{-0.25cm}
\end{equation}%
where $Y_{it}$ is log-sales or IHS-profits\footnote{IHS stands for \textit{inverse hyperbolic sine} and is calculated as $L(x)=\log(x+\sqrt{x^2+1})$ following \cite{mackinnon1990transforming}. It is preferable to log transformations for variables containing negative values.} of firm $i$ at year $t$; $\alpha_i $ and $ \delta_t $ are firm and year fixed effects; and $ \Delta \widehat{\textnormal{Centrality}}_{i,2013} $ is the exogeneous war-induced change in centrality after the start of the conflict.
%The coefficients of interest are the year-by-year coefficients $\beta_{t}$ on the predicted change in firm centrality. 
Under the parallel-trends assumption, $\beta_t$'s identify the impact of an exogeneous change in a firm's network position postconflict on the firm's subsequent performance.

We use three measures of centrality: eigenvector centrality, betweenness centrality, and degree centrality, and report the results for each of the measures in Figure~\ref{FIG: Sales and Centrality 1}.\footnote{We present the exact definitions of these centrality measures in Appendix B.} For all centrality measures, the effects take off in 2014, are persistent, and are generally increasing over time. Before the conflict, firms with different values of predicted centrality evolved along the same trends, which gives credibility to the difference-in-differences design. According to Table~\ref{TABLE: Centrality}, which presents the estimates of a version of equation~\eqref{eq: centrality} replacing $\beta_t$ with a simple pre-post $\beta$, a one-standard-deviation larger exogeneous change in centrality leads to an 8.9\% to 14.5\% increase in firm's sales.

%Figure 4
\begin{figure}[!t]
\begin{center}
\begin{minipage}{0.95\textwidth}
    \begin{center}
        \includegraphics[width=\textwidth]{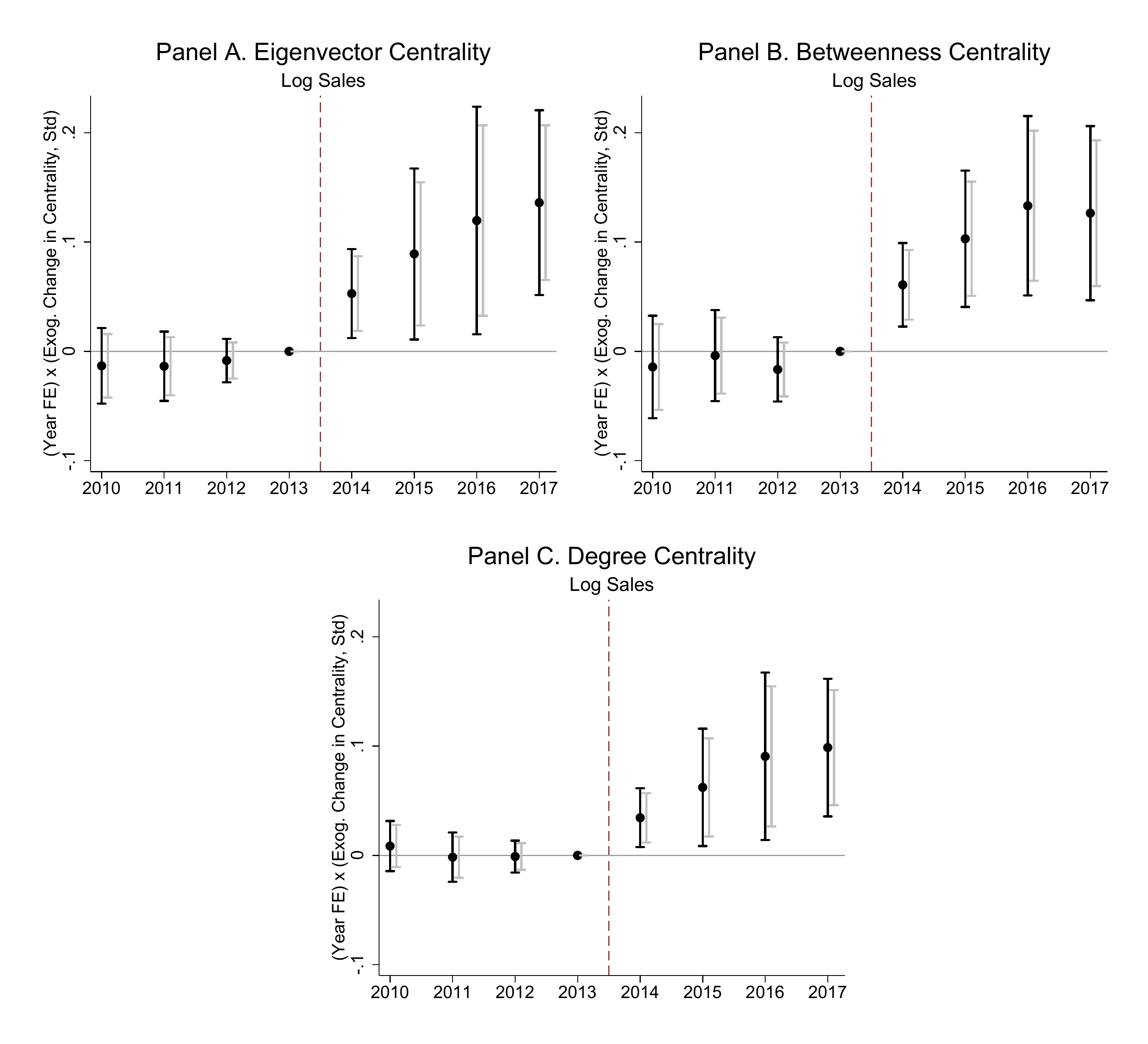}
    \end{center}
\caption{Conflict-Induced Change in Network Centrality and Sales}\label{FIG: Sales and Centrality 1} %
%\vspace{-0.2cm}
\renewcommand{\baselinestretch}{0.7}
\footnotesize{\textit{Notes}: This figure displays the results of estimating equation \eqref{eq: centrality}. Panel A displays the results for the predicted change in eigenvector centrality between 2013 and 2014 as the interaction variable, Panel B displays the results for the predicted change in log betweenness centrality between 2013 and 2014, and Panel C displays the results for the predicted change in degree centrality between 2013 and 2014. Black bars represent 95\% confidence intervals, gray bars represent 90\% confidence intervals. All measures of centrality changes are standardized to have zero mean and a standard deviation of one. The sample excludes firms located in the conflict areas. Standard errors are clustered at the firm level. $N=27,201$.}
\vspace{-0.3cm}
\end{minipage}
\end{center}
\end{figure}

\begin{figure}[!t]
\begin{center}
\begin{minipage}{0.95\textwidth}
    \begin{center}
        \includegraphics[width=\textwidth]{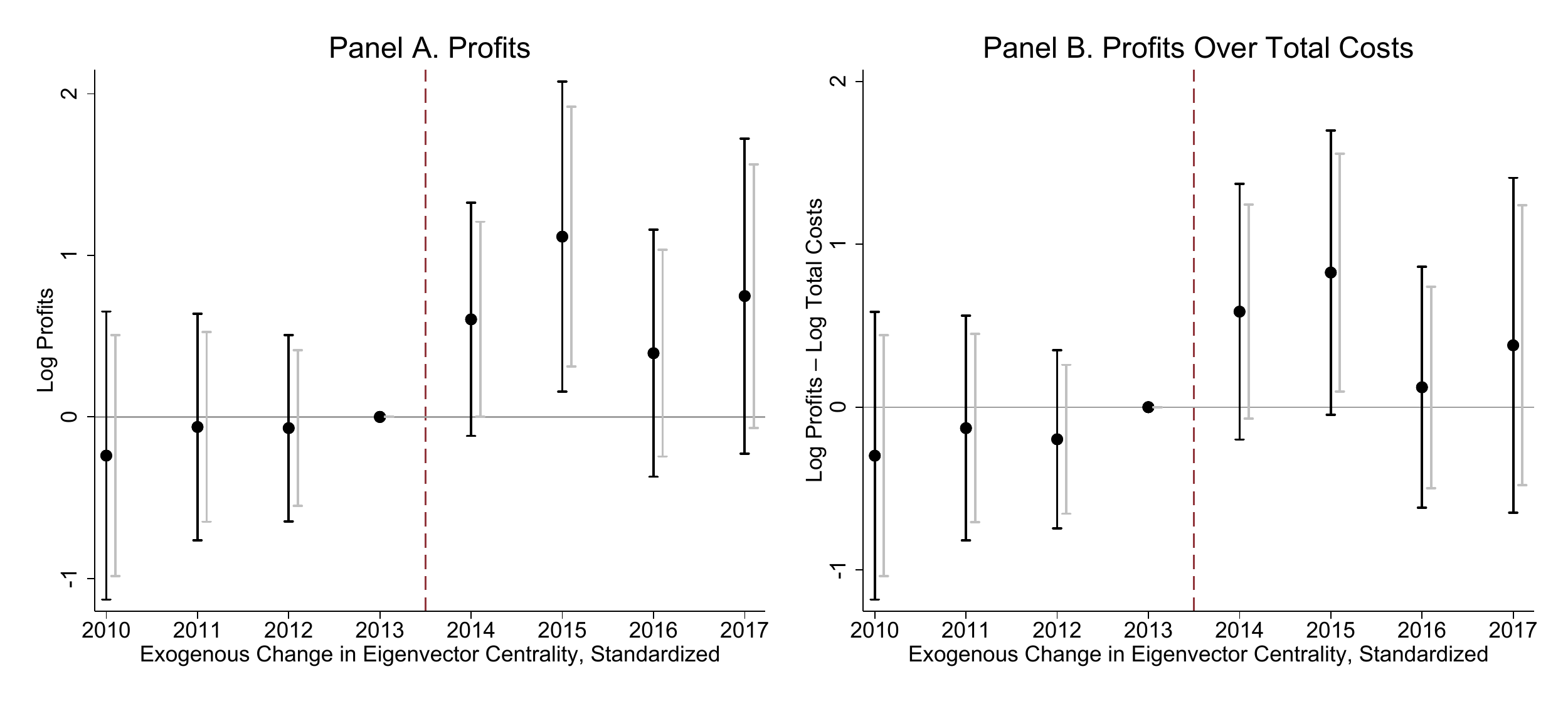}
    \end{center}
    %\vspace{-0.2cm}
\caption{Conflict-Induced Change in Network Centrality, Profits, and Profit-to-Cost Ratio}\label{FIG: Profits and Markups} %
%\vspace{-0.2cm}
\renewcommand{\baselinestretch}{0.7}
\footnotesize{\textit{Notes}: This figure displays the results of estimating equation \eqref{eq: centrality}. Both panels display the results for the predicted change in eigenvector centrality between 2013 and 2014 as the interaction variable, standardized to have zero mean and standard deviation of one. Panel~A presents the results for the inverse hyperbole sine of gross profits as the dependent variable. Panel~B presents the results for the inverse hyperbole sine of profits minus the inverse hyperbole sine of total costs as the dependent variable. Black bars represent 95\% confidence intervals, gray bars represent 90\% confidence intervals. Standard errors are clustered at the firm level. }
\vspace{-0.3cm}
\end{minipage}
\end{center}
\end{figure}

Next, Figure~\ref{FIG: Profits and Markups} presents the effects on profits and the profits-over-costs ratio. %
%\footnote{Ukrainian and Russian accounting systems provide annual total costs of goods and services produced. We evaluate markups based on this measure. Specifically the main outcome variable is the difference between the log-gross profits and log-total costs. We transform both variables using an IHS transformation to deal with zero values.} %
The results are in line with a story in which, in the short-run, firms that become more central charge higher prices which, in turn, translate into higher profits. Both effects dissipate by the end of the study period. 
%At the same time, the shift in centrality has a longer-term consequences for sales, since sales shift in the medium- to long-run as well.

%Figure 6

One may wonder if the results displayed in Figure~\ref{FIG: Sales and Centrality 1} are simply a mirror image of the propagation effects on trade presented earlier in Section~\ref{sec:results_propagation}. That is, it could be that firms that traded with the conflict areas by definition became less central with the start of the conflict. To assuage this concern, we present two additional exercises. First, we directly control for a firm's ties with the conflict areas at baseline interacted with the yearly indicators.\footnote{Specifically, we estimate the following specification:
\vspace*{-0.25cm}
\begin{gather} \label{eq: joint}
		Y_{it} = \alpha_i + \delta_t + \beta_{t} \times \Delta \widehat{\textnormal{Centrality}}_{i,2013} + \gamma_{t} \times \textnormal{Conflict}_{i,2013} + \varepsilon_{it}
\vspace*{-0.25cm}
\end{gather}%
where $Y_{it}$ is either log-sales or IHS-profits of firm $i$ at year $t$; $\alpha_i $ and $ \delta_t $ are the firm and year fixed effects; $ \Delta \widehat{\textnormal{Centrality}}_{i,2013} $ is the exogeneous conflict-induced change in centrality after the start of the conflict; and $\textnormal{Conflict}_{i,2013}$ is an indicator for whether $i$ traded with the conflict areas in 2013.} 
Figure \ref{FIG: Sales, Network Centrality and Conflict} plots the estimates. Panel A presents the centrality coefficients $\hat \beta_t$, which appear to be virtually unchanged compared to Figure \ref{FIG: Sales and Centrality 1}. Panel~B reports the marginal negative impact of having conflict ties, one of the mechanisms for which could be the propagation effects on trade documented in Section~\ref{sec:results_propagation}. 

Second, we further account for propagation with a residualization exercise.
Specifically, we residualize the exogeneous change in centrality from firms' baseline characteristics, such as whether a firm traded with someone in the conflict areas, the share of its transactions being with the conflict areas, the share of its total sales being with the conflict areas, the total weight of its partners' trade with the conflict areas (i.e., through second-degree connections), as well as the firm's baseline sales and profits. We then use the obtained residuals as a measure of exogenous change in centrality in equation~\eqref{eq: centrality}. The results presented in Figure~\ref{FIG: Residualizing Change in Centrality} further confirm that the results in Figure~\ref{FIG: Sales and Centrality 1} remain robust to controlling for indirect exposure to conflict.  %Some of the coefficients in Panel B are marginally insignificant; however, overall, there was a sizeable drop in sales for firms that had ties with the conflict areas in 2013. Specification~\eqref{eq: joint} allows us to separate the first-degree effects of conflict ties and the effects of exogenous change in centrality.

Overall, these results detail a previously undocumented indirect effect of conflict, and possibly other large persistent shocks cutting off certain regions from trade with the rest of the country, on firms through a change in the production-network structure. We confirm such effect's existence by showing that Ukrainian firms which, for reasons beyond their control, became more central with the start of the Russia-Ukraine conflict gained (or lost less) in terms of sales and profits. We speculate that these effects may be due to a temporary surge in those firms' relative market power.

%\paragraph{Back-of-the-envelope calculations.} 
%We can use the latter results and combine the centrality effects and the effects coming from ties to conflict to understand the aggregate effect outside of the conflict zone in a rough back-of-the-envelope calculation. We note that 37\% of the firms traded with conflict areas before the conflict. The average treatment effect from a pooled version of~\eqref{eq: joint} is $-$11.9\%, and thus a combined effect of conflict ties on sales is $-$4.4\%. The average treatment effect for betweenness centrality is 6.4\%, with an average shift in centrality of $-$0.22, and a corresponding aggregate effect of $-$1.4\%. Summing up the conflict effect and the centrality effect, one gets at the $-$5.8\% of the total change.

%\section{Channels}
%\label{sec:channels}

\section{War, Network Destruction, and Network Adjustment}
\label{sec:concframework}

So far, we have documented that a large-scale secessionist conflict (i) disrupts trade, and this disruption propagates even to trade links in nonconflict areas, and (ii) alters the structure of the production network, which on its own affects firm performance. In this section, we go a step beyond relative comparisons and use a theoretical model to explore the war-induced absolute changes to revenue distribution in several counterfactual scenarios. Most crucially, we quantify the importance of endogenous network formation for economic recovery from a large negative shock.
%Specifically, we employ model-based counterfactuals to quantify the three main channels (network destruction, network adjustment, and a shock to outside demand) through which war affects the production network and, ultimately, the distribution of firm revenue.

\subsection{Model}
\label{sec:model}

We build a general equilibrium model of production networks in the spirit of \cite{long1983real}, \cite{acemoglu2012network}, and \cite{baqaee2019macroeconomic}. The model allows us to explore three counterfactuals: (i) the first one corresponds to network destruction before firms manage to adjust to the new equilibrium and redirect their input and output flows; (ii) the second one corresponds to the observed network adjustment of the firms; (iii) the third counterfactual focuses only on the outside demand changes after the conflict, keeping the production network fixed at the preconflict level.

To accomodate these counterfactuals, we depart from the standard Cobb-Douglas production function (e.g., \citealp{acemoglu2012network}) in favor of a more general constant elasticity of substitution (CES) technology.%
\footnote{Our empirical exercise is different from the existing literature (\citealp{acemoglu2012network}, \citealp{carvalho2021supply}) in one critical aspect: we directly observe interfirm trade flows and, thus, firm-level input-output matrices both before and after the shock, which allows for studying more general counterfactuals.} %
Specifically, in our hypothetical economy, a generic firm $i$ uses a bundle of intermediate inputs described by a CES aggregator: $M_i=\Big{(} \sum_{j=1}^n a_{ji}^{1/\sigma}x_{ji}^{(\sigma-1)/\sigma} \Big{)}^{\frac{\sigma}{\sigma-1}}$ where $x_{ji}$ is the amount of good $j$ used in production of good $i$ and $a_{ji}$ is the share of input $j$ needed for good~$i$. We assume constant returns to scale for intermediate inputs; that is, $\sum_{j=1}^n a_{ji} = 1$.\footnote{This assumption is to simplify the derivations; however, one can consider a more general case of decreasing returns to scale with $\sum_{j=1}^n a_{ji} < 1$.}

The substitution patterns between the CES intermediate-input aggregator, $M_i$, and the only other factor, labor $l_i$, take a Cobb-Douglas form with constant returns to scale. In this case, output $x_i$ of firm $i$ can be written as:%
%\footnote{Note that the standard Cobb-Douglas formulation used in, e.g., \cite{acemoglu2012network} can be derived if the elasticity of substitution between intermediate input bundle $M_i$ and labor tends to unity. The CES form is much more helpful in our context since Cobb-Douglas would predict zero output if a firm could not source one of the necessary inputs. With the CES, firms can restructure their inputs and keep producing even when one of the preconflict inputs is zero.}
\vspace*{-0.25cm}
\begin{gather*}
    x_i = (z_i l_i)^{\alpha} \Big{(} \sum_{j=1}^n a_{ji}^{1/\sigma}x_{ji}^{(\sigma-1)/\sigma} \Big{)}^{\frac{\sigma}{\sigma-1}(1-\alpha)}
\vspace*{-0.25cm}
\end{gather*}
where $z_i$ is firm-specific productivity, $l_i$ is the amount of labor hired, and $x_{ji}$ is the input of good $j$ used in the production of output $i$. 

We assume no free entry, which corresponds to a short- or medium-run case, and firms earn positive profits in equilibrium with constant markups.\footnote{We do not explicitly model entry throughout the theoretical part. Incorporating entry in an exogenous way is relatively straightforward; see \cite{melitz2003impact}, \cite{chaney2008distorted} or \cite{magerman2016heterogeneous} for examples. As long as firms base their decisions to enter only on their productivity or costs draws, and conflict happens in an unrelated way---which is a natural assumption---it should not affect the derivation of the main equation that ties revenues, outside demand, and the input-output linkages in our economy. We detail how we treat actual entrants and leavers in the empirical part of this section. With free entry, firms can enter until profits are equal to the entry costs.}

We combine the CES production technology with CES preferences over $n$ goods, so the consumer maximizes the following utility function:
\vspace*{-0.25cm}
\begin{gather*}
    u(c_1,...,c_n) = \Big{(} \sum_{i=1}^n c_i^{\eta} \Big{)}^{1/\eta}
\vspace*{-0.25cm}
\end{gather*}
The constant across-goods elasticity of substitution is $1/(1-\eta)$ in this case.

The supply of labor is inelastic, with the total labor in the economy normalized to $L$. Wages are denoted by $w$.

A competitive equilibrium of our railway economy with $n$ firms and $n$ differentiated goods consists of the price vector $\mathbf{p}$, wage $w$, consumption vector $\textbf{c}$, and quantities $(l_i, x_i, \mathbf{x_{ji}})$ such that (i) the representative consumer maximizes utility; (ii) representative firms maximize their profits given the prices; (iii) all markets clear:
\vspace*{-0.25cm}
\begin{gather*}
    x_i = \sum_{j=1}^n x_{ij} + c_i,\ \forall\ i,
    \sum_{i=1}^n l_i = L
\vspace*{-0.25cm}
\end{gather*}
Expressing $x_{ij}$ from the first-order condition of the firms' profit-maximization problem (see the derivations in Appendix C), we obtain:
\vspace*{-0.25cm}
\begin{gather*}
  r_i = (1-\alpha) \sum_{j=1}^n \underbrace{\frac{a_{ij}/p_i^{1-\sigma}(\mathbf{a})}{\sum_m a_{mj}/p_m^{1-\sigma}(\mathbf{a})}}_{\text{I-O weight }\omega_{ij}(\mathbf{a})}  \cdot r_j + \underbrace{p_i c_i}_{\text{outside demand } \bm{\xi}}
\vspace*{-0.25cm}
\end{gather*}
We can summarize the input-output linkages with matrix $\mathbf{\Omega}$, which simultaneously represents two things: the adjacency matrix observed in the data, and the economy's steady-state input-output matrix. Concise matrix notation allows us to rewrite the central equation as one relating the vector of revenues, outside demand, and the input-output matrix in the economy:
\vspace*{-0.25cm}
\begin{equation} \label{eq: leontief_raw}
    \mathbf{r}=(1-\alpha) \mathbf{\Omega} \mathbf{r} + \bm{\xi}
\vspace*{-0.25cm}
\end{equation}
Note that $\mathbf{\Omega}$ and $\mathbf{r}$ are observed in the data, and each element of the matrix is defined by the equilibrium weights, $\omega_{ij}(\mathbf{a}) = \frac{a_{ij}/p_i^{1-\sigma}(\mathbf{a})}{\sum_m a_{mj}/p_m^{1-\sigma}(\mathbf{a})}$. Vector $\bm{\xi}$ is the outside demand vector, which includes both consumer demand and demand from firms outside the railway network.

If $\sigma\rightarrow1$, our production function converges to Cobb-Douglas and elements of $\mathbf{\Omega}$ ($\omega_{ij}$) converge to constants proportional to $a_{ij}$. The main difference between our CES aggregator and the Cobb-Douglas is that zero inputs for some intermediate goods does not automatically mean zero production for the firms. However, with the CES production function, solving for the equilibrium in a closed form becomes impossible. Our empirical exercise overcomes the intractability of the model by manipulating $\mathbf{\Omega}$ directly. 

 \subsection{Counterfactuals}

For the empirical part of this exercise, we add yearly indices $t$ to the input-output matrix equation~\eqref{eq: leontief_raw} to get:\footnote{Adding yearly indices does not change the substance of the model. Their first purpose is to stress that if the inputs of the model---the outside demand---change, so do the outputs, i.e., the Leontief matrix and the revenue vector. The second purpose of adding them is notational convenience. While the model is static, we think the yearly data correspond to the new equilibrium if there is a shock.}
\begin{equation} \label{eq: leontief}
        [\mathbf{I} - (1-\alpha^t) \mathbf{\Omega}^t] \mathbf{r}^t = \bm{\xi}^t
%\vspace*{-0.25cm}
\end{equation}
where $\bold{I}$ is an identity matrix. Following the notation in Section~\ref{sec:model}, $\bold{\Omega}^t$ is the railway firm-level input-output matrix;
%Our matrix boils down to exactly the cost-based matrix from \cite{baqaee2020productivity} if we assume that prices of inputs are constant across suppliers of the same firm. This assumption is much more plausible than assuming that input and output prices are fixed for the same firm, the one we would have needed if we wanted to link $\mathbf{\Omega}$ to the revenue-based input-output matrix.}
$\bold{r}^t$ is a vector of firms' total revenues in year $t$; $\alpha^t$ is the labor share in production (or any non-intermediate-factor share in our case) in year $t$; and $\bm{\xi}^t$ is yearly outside demand. We fix labor share to be constant in time $\alpha^t=\alpha$ and allow revenues, final demand, and the input-output matrix to vary flexibly over $t$. We then plug in $\bold{r}^t$ and $\bold{\Omega}^t$ in order to back out the final demand $ \bm{\xi}^t$.\footnote{The choice of $\alpha$ is potentially important. We estimate $\alpha$ from the accounting data, using the 2013--2015 sample of firms and a simple log-log regression with fixed effects. The resulting $\hat \alpha = 0.18$ can be interpreted as a weighted yearly average of $\alpha^t$. This estimate is quite similar to the \cite{magerman2016heterogeneous} number for Belgium, where $\alpha = 0.19$. An alternative approach is to directly use the official input-output tables for Ukraine, where $\hat \alpha$ varies between 0.15 and 0.2 depending on the year.} One key advantage of our data is that we observe the realized $\bold{\Omega}^t$ and thus can vary it to explore different counterfactuals relating production-network structure and firm revenue.\footnote{We rely on total weight shipped as a measure of quantity in our baseline analysis, which makes our input-output matrix $\mathbf{\Omega}$ similar to the \textit{cost-based} rather than the \textit{revenue-based} input-output matrix from \cite{baqaee2020productivity}, with the former being the one advocated by the authors. Figure~\ref{FIG: Weight Value Correlation} shows that total weight and total value are highly correlated in the Ukrainian customs data, further alleviating the concern of potential mismeasurement.}  
%Namely, we can construct a predicted change of $\bold{\Omega}^t$ from 2013 to 2014 by cutting the links with the conflict areas entirely, similarly to what we did in Section \ref{sec:results_centrality}, and compare it to the actual $\bold{\Omega}^t$ in 2014.

%Our ultimate goal is to contrast the network destruction channel's influence size and the size of the network adjustment, keeping outside demand fixed at the 2013 level. As is typical in the production-network literature---see, for example, \cite{baqaee2020productivity} and, closer to our setting see \cite{magerman2016heterogeneous}, we do not address the exit and entry of firms directly.\footnote{See \cite{baqaee2018cascading} for details of a model of extensive margin of firm entry and exit in a production network.} Instead, we explore robustness to different ways of dealing with firms that drop from the data. In our baseline exercise, we keep the firms from 2013 and impute zero revenues to firms with missing revenues later on.\footnote{See Panel B of Table \ref{TABLE:counter_expanded} in the Appendix for an alternative approach where we keep all of the firms who appeared in the data at least once, and impute zeros for the missing revenues.}

The first counterfactual we explore gets at the aggregate impact of the initial shock of war to the production-network structure. It corresponds to the change from $\bold{\Omega^{2013}}$ to $\bold{\tilde \Omega^{2013}}$, which is the 2013 network but with zeros in place of trade with the firms in the conflict areas (object similar to the one used in Section~\ref{sec:results_centrality}).\footnote{We impute zeroes as an approximation for the conflict practically cutting off DPR, LPR, and Crimea from the rest of Ukraine. Figure~\ref{FIG: Conflict Production Network Weights} shows that this approximation is not far from reality. The graph displays the quartiles of the production network weights $\omega_{ij}$ corresponding to trade with the conflict areas. All quartiles decrease sharply in 2014 and reach precise zeroes in 2015 and 2016. Non-zero trade in 2014 likely occurred during the peaceful months, such as January 2014.} Final demand is kept constant at $\bm{\xi^{2013}}$. This counterfactual quantifies the \textit{network-destruction} channel.\footnote{One might ask how this first counterfactual maps into our model where returns to scale are constant and the input shares are determined in equilibrium (the inputs are fixed in the limiting case of Cobb-Douglas). We think about network destruction as a short-term scenario in an environment where firms were bound by contracts with fixed prices and quantities, and the upstream firms were not able to deliver some of the prespecified input quantities in 2014. From the model's point of view, labor compensates for some of the missing inputs. Note that we keep the labor share fixed in our baseline empirical exercise. Instead, if we reestimate the labor share from the accounting data for each year and use those numbers, the results remain very similar to our baseline case. Then, in the network-adjustment counterfactual, we use the actual 2014 production network that can be thought of as allowing contracts to be flexible.} 

The second counterfactual has to do with the economy's adjustment. Specifically, after the initial shock, firms may have found new buyers and suppliers within the nonconflict areas, and this readjustment could have mitigated the initial impact of the war. To reflect this change, we keep the final demand at the 2013 level, $\bm{\xi^{2013}}$, and we plug in the new network structure from 2014, $\bold{\Omega^{2014}}$. The difference between the resulting change and the change in the network-destruction counterfactual quantifies the \textit{network-adjustment} channel.

Finally, to be transparent about all of the forces that drive the change before and after conflict, we also consider the \textit{outside-demand} counterfactual. Specifically, we assess the size of the outside demand shock by keeping the railway production network at $\bold{\Omega}^{2013}$ and varying the final demand from $\bm{\xi^{2013}}$ to $\bm{\xi^{2014}}$. To estimate $\bm{\xi^{2013}}$ and $\bm{\xi^{2014}}$, we plug in $(\bold{\Omega^t}, \bold{r}^t),\ t=\{2013,2014\}$.

Table \ref{table:counterfactuals} presents the results for each counterfactual. Column~2 of Table \ref{table:counterfactuals} reports the change in the median of the distribution, corresponding to the network-destruction channel. Network destruction leads to a substantial $46.8\%$ drop in the median firm revenue in nonconflict areas.\footnote{In our baseline exercise, we only keep the firms from 2013. See Table~\ref{TABLE:counter_sample} for an alternative approach, where we keep all of the firms that appear in the data at least once and impute zeros for the missing revenues. The results remain qualitatively similar.} 

Column 3 of Table \ref{table:counterfactuals} suggests that, if the production network were allowed to readjust and firms were allowed to find new partners, the initial shock to the network structure manifested by the changes in $\bold{\Omega}^t$ would be mitigated substantially. Network adjustment compensates for around 80\% of the network destruction for the firm median revenue in nonconflict areas, with a total decline of $-$9.7\% relative to the baseline. 

Column 4 of Table \ref{table:counterfactuals} displays the results for the outside-demand counterfactual. The demand shock is equivalent to a $17.3\%$ decline in the median of the revenue distribution, i.e., smaller than network destruction in magnitude but larger than if firms are allowed to adjust their network.

Finally, we can compare all of these channels to the total observed change, which is $-27.2\%$ relative to the 2013 baseline (column~5). The actual change in this case turns out to be a rough sum of the network-adjustment and outside-demand effects.
%closer to the network-destruction counterfactual, that is, to the case when in the short term firms cannot quickly substitute their inputs and either source more from existing nonconflict suppliers, or find new suppliers.

To summarize the above, the exogenous change in production-network structure due to conflict not only creates relative winners and losers, as in Section~\ref{sec:results_centrality}, but also moves the distribution of firm revenue down, decreasing the median firm revenue in the rest of the country by 46.8\%. However, readjustment after the start of the conflict within the remaining network compensates four-fifths of the latter decline, suggesting that endogenous network formation compensates for most of the initial shock. The following section explores the dynamics of these two effects, considers changes along the whole distribution, and extends the counterfactuals to all firms in the economy in an aggregation exercise.

\begin{table}[!]
\caption{Counterfactuals}\label{table:counterfactuals}

\resizebox{0.95\textwidth}{!}{\begin{tabular}{lccccc}
\hline
 & (1) & (2) & (3) & (4) & (5) \\

& $\bold{r}(\bold{\Omega^{2013}}, \bm{\xi^{2013}})$ & $\bold{r}(\bold{\tilde \Omega^{2013}},\bm{\xi^{2013}})$ & $\bold{r}(\bold{\Omega^{2014}},\bm{\xi^{2013}})$
 &  $\bold{r}(\bold{\Omega^{2013}},\bm{\xi^{2014}})$ & $\bold{r}(\bold{\Omega^{2014}},\bm{\xi^{2014}})$   \\

\hline

& \makecell{Baseline \\ \ } & \makecell{Network \\ Destruction} & \makecell{Network \\ Adjustment}  & \makecell{Outside \\ Demand} & \makecell{Total \\ Change}    \\ \hline

% Median $\bold{r}$, Mln. Hr. & 17.9 & 12.1 & 15.9 & 15.0 & 13.0 \\ 

% Relative to 2013 & 0.0\% & -32.4\% & -11.0\% & -16.0\% & -27.2\% \\ 

$\text{Median}(\bold{r})$, Mln. 2010 US\$ & 2.26 & 1.20 & 2.03  & 1.86 & 1.64  \\ 

Relative to 2013 & 0\% & -46.8\% & -9.7\% & -17.3\% & -27.2\%  \\
%\hline

$N$ & 4,463 & 4,463 & 4,463 & 4,463 & 4,463 \\

\hline

\hline

\end{tabular}
}

\begin{minipage}{\textwidth}
\vspace{0.2cm}
\renewcommand{\baselinestretch}{0.7}
\footnotesize{\textit{Notes:} The table presents the counterfactual estimates. Specifically, we obtain the vector of revenues $\bm{r}$ for a given input-output matrix $\bold{\Omega}$ and outside demand $\bm{\xi}$ such that    $\bm{r}(\bm{\xi},\bold{\Omega}):=[\bold{I}-(1-\alpha)\bold{\Omega}]^{-1}\bm{\xi}$, and we present the median revenue of the resulting distributions. Columns vary each of the arguments. Column (1) displays the actual median firm revenue in 2013. Column (2) uses the modified 2013 input-output matrix, where links with firms from the conflict areas are replaced with zeros, $\bold{\tilde \Omega^{2013}}$. Column (3) fixes the outside demand at the 2013 level and uses the actual 2014 input-output matrix, $\bold{\Omega^{2014}}$. Column (4) uses the 2014 outside demand and the 2013 input-output matrix, $\bold{\Omega^{2013}}$. Column (5) displays the actual median firm revenue in 2014. The sample is restricted to firms in nonconflict areas. All quantities are expressed in 2010 US\$.
}
\vspace{-0.3cm}
\end{minipage}
\end{table}

%%%%%%%%
%%%%%%%%
%%%%%%%%%
%%%%%%%%%

\subsection{Additional Results: Interfirm Inequality and Dynamics of Network Adjustment}

\paragraph{Interfirm inequality.} \label{sec:interfirm_inequality}

While the change in the median is a crucial summary statistic for our counterfactuals, one might also ask what happens to other parts of the firm revenue distribution. As it turns out, the initial conflict shock and the subsequent network adjustment led to an increase in inequality between firms. To document it, we extend the estimates in Table~\ref{table:counterfactuals} to other features of the firm-revenue distribution and compare the network-adjustment counterfactual to the network-destruction scenario and the observed data for firms at the left and right tails.

Table \ref{TABLE:counter_expanded} presents the results. Column~3 of Table \ref{TABLE:counter_expanded} suggests that firms at the 75th percentile of the distributions \textit{gain} 14.0\% of their 2013 revenues in the network-adjustment counterfactual---compared to both the observed loss of 22.8\% in column~5 and the loss of 29.3\% in the network-destruction counterfactual in column~2. Hence, we find that, at the right tail of the distribution, network adjustment fully compensates for the loss of revenue due to network destruction and even leads to some gains. In contrast, at the 25th percentile of the distribution, firms lose slightly more under the network-adjustment scenario than in the observed data---a 62.3\% vs.\ 62.0\% drop. Moreover, network adjustment compensates for only 32.5\% of the network destruction at the left tail---much less than at the median. Overall, the estimates imply that, controlling for outside shocks and other factors, large adverse shocks to portions of the network and subsequent network adjustment exacerbate interfirm inequality and may lead to greater concentration.

\vspace*{0.5cm}

\noindent \textbf{Dynamics.} \label{sec:dynamics} %
Another natural question to ask is whether network adjusted immediately or whether this process stretched over time, e.g., due to switching costs, and thus compensated for more and more of the initial network-destruction shock as time went on. 

\begin{figure}[!t]
\begin{center}
\begin{minipage}{0.9\textwidth}
    \begin{center}
        \includegraphics[width=0.9\textwidth]{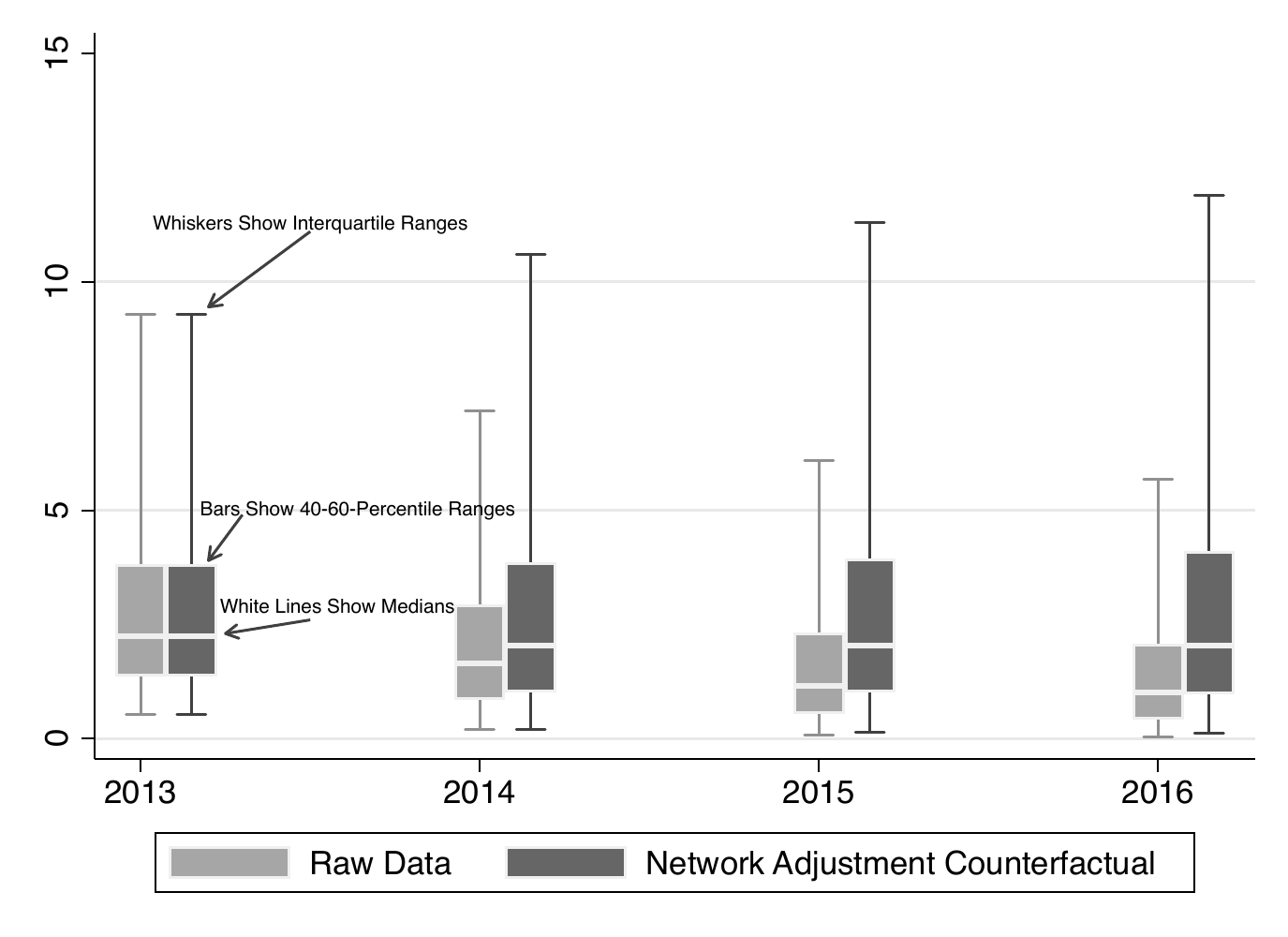}
    \end{center}
\caption{Network-Adjustment Counterfactual and Actual Changes Over Time}\label{FIG: NA over time} %
\renewcommand{\baselinestretch}{0.7}
\footnotesize{\textit{Notes}: The graph depicts the 2013--2016 evolution of two firm revenue distributions: the one we actually observe in the data (in light grey) and the one we derive under the network-adjustment counterfactual scenario (in dark grey). The ``whiskers'' represent the interquartile range, and the bars represent the 40th-to-60th-percentile range of each distribution. The network-adjustment counterfactual scenario is derived by fixing the outside demand at the 2013 level, $\bm{\xi}^{2013}$, and changing the input-output matrix gradually, $\bold{\Omega}^t:\ t=\{2013,2014,2015,2016\}$. All quantities are expressed in 2010 \$US. $N$=4,463 per year. The sample is a balanced panel of the 2013 firms.}
\end{minipage}
\end{center}
\end{figure}

To study the medium-run effects of the network adjustment to conflict, Figure \ref{FIG: NA over time} plots the medians and the interquartile ranges of 2013--2016 firm revenue. We find that, under the network-adjustment counterfactual, median revenue did not vary across years: it initially changed by $-$9.7\% in 2014, then stayed low at $-$9.9\% in 2015 and $-$9.4\% in 2016 (all relative to the 2013 median). 
%Thus, at least at the median, the production network did not adjust significantly more over time. 
%but full compensation did not occur in three years after the shock.

% To study the medium-run effects of the network adjustment to conflict, we plot the medians and the interquartile ranges of firm revenue for 2013--2016 in Figure \ref{FIG: NA over time}. First, we find that, under the network-adjustment counterfactual, median revenue initially changes by $-9.7\%$ in 2014, then stays low at $-9.9\%$ in 2015 relative to the 2013 median, and almost fully recovers by 2016 ($-1.0\%$ relative to 2013 and $+11.2\%$ relative to 2014). Thus, the production network adjusts rather quickly but full compensation occurs only two to three years after the shock.

Instead, we find substantial dynamics at the tails of the distribution. Specifically, the 75th percentile under network adjustment grew by 14.0\% in 2014, 21.4\% in 2015, and by 28.0\% in 2016 relative to the same point in the revenue distribution in 2013. At the same time, the 25th percentile declined by $-$62.3\% in 2014, $-$73.0\% in 2015, and by $-$77.5\% in 2016. The 40th-to-60th-percentile ranges in Figure~\ref{FIG: NA over time} further confirm this pattern. 

Thus, the endogeneous network-adjustment channel, controlling for outside demand and other factors, pushes toward increased interfirm inequality and concentration over time, with little growth at the median after the initial adjustment.

%Overall, our findings suggest that the network adjustment allowed the economy to mitigate the negative network-structure shock at the median with little growth over time. Nevertheless, the growth took place at the right tail of the distribution, with inequality between firms growing in the medium run due to this channel.

% \vspace*{0.5cm}

\subsection{Extrapolating Beyond the Railway Network of Firms} \label{sec:agg_level}

At the final step of analysis, we address the question of external validity of our results in the third part of the paper outside the railway network of firms. To do so, we aggregate our estimates to the near universe of all firms in the economy\footnote{There are 1,334,052 firm-year observations for 2013-2016 in our accounting data. Nevertheless, the 4,700 firms in our baseline railway sample account for 29\% of all revenues in the economy.} and compare the overall change and the network-adjustment counterfactual at a higher geographical level. 

To do so, we redefine the vector of revenues and the matrix of connections at the province and district levels. We take the total of all firm revenues for a given geographic unit and year as the $\bold{r}$ input. We proxy for the aggregate-level $\bold{\Omega}$ using railway data by summing up all transactions for each pair of localities.\footnote{Note that, in these calculations, many rayons go omitted because they do not have railway connections.} The disadvantage of this exercise is that it assumes that the measurement error of connections is similar across locality pairs. However, it extends the external validity of our results by focusing on the near-universe of firms in the economy rather than only on firms relying on the rail transport. 

We assess the results of this aggregate counterfactual estimation in two steps. First, we plot the province- and district-level maps in Figures~\ref{FIG: NA Maps Province} and~\ref{FIG: NA Maps District}, respectively, with the top maps in both figures representing the observed changes and the bottom maps representing the network-adjustment counterfactual changes. At both aggregation levels, network adjustment smoothens the changes compared to the observed data. What is especially evident from the figures, particularly for the more granular depiction at the district level, is that areas near but just outside the conflict zones take a big hit in the observed data  but can benefit under the network-adjustment scenario. This finding highlights the mechanics of our network-adjustment exercise: it controls for outside demand shocks but allows for reallocation of trade within the rest of the country, with nearby districts seemingly serving as a good substitute for production in the conflict areas.
 
Second, in Table \ref{table:regions}, we collapse these geographic patterns into one number---countrywide total revenue. Total revenues coming from the province-level estimates in columns~(2) and~(3) and the district-level estimates in columns~(4) and~(5) are largely similar. They exhibit a deterioration in the raw data but a gradual increase in the adjustment counterfactual. Overall, these results are in line with the intuition that network adjustment represents a tool for economic recovery.%
\footnote{The absolute changes for network adjustment in Table \ref{table:regions} relative to the 2013 baseline are positive while they are negative in our previous subsection. Here, we look at changing the production structure fixing demand---this allows some firms to produce more, but also creates winners and losers. While this channel yields a negative effect for a median firm, shutting down all other channels at a higher geographic level and then aggregating the results to a countrywide total yields a positive effect.}

\section{Conclusion}
\label{sec:conclusion}

Over the past few decades, the world has become exceedingly interconnected, and production networks have grown ever more complex. On one hand, due to increased complexity, economic systems have become vulnerable to large shocks. On the other hand, any such shocks are now easier to absorb due to an increased number of potential substitutes. Given these worldwide trends, it is now increasingly important to study how modern production networks react to big shocks, such as wars or pandemics, how quickly they are able to readjust, and which firms stand to gain or lose as a result. Despite recent progress in this research agenda, evidence remains limited, both in terms of research questions and its regional scope.

In this paper, we examine the consequences of a particular massive shock to the economy---large-scale conflict. Specifically, we study the ramifications of the 2014 Donbas war and the annexation of Crimea for the Ukrainian economy. Using novel data on the universe of Ukrainian firm-to-firm railway shipments in 2013--2016, we document that the conflict shock did indeed propagate throughout the production network and spill over onto the rest of the country. Trade has significantly decreased, even between firms outside of the conflict areas, if one of them had a trade partner in Donbas or Crimea before the start of the conflict. 

We then study the consequences of the shock to the production-network structure by which firms in three large regions became virtually cut off from the rest of the economy. We find that firms which, for exogenous reasons, became more central have experienced significant permanent gains in sales. These are accompanied by temporary gains in profits and the profits-over-total-costs ratio, suggesting increased market power as a plausible mechanism. Finally, estimating counterfactuals in a production-network model, we show that production-network adjustment would be able to mitigate 80\% of the losses in median firm revenue after the initial network-destruction shock, keeping outside factors the same. %Nevertheless, full readjustment takes time, up to two to three years in this particular context.

Overall, our results suggest that large shocks not only propagate through the network but also may alter its structure, with potentially significant implications for the economy in the medium run. Our results should be of interest to academics and policymakers interested in estimating, predicting, and mitigating the economic consequences of conflicts and other large economic shocks.

Despite the numerous differences between the 2014 Russia-Ukraine conflict and the full-scale Russian invasion of Ukraine in February 2022, our estimates offer a unique glimpse into the potential economic consequences of the ongoing conflict that have not yet been discussed in the academic literature. We hope that future research will shed more light on the full ramifications of the latter episode. 
%However, further evidence is needed to better understand the mechanisms behind our estimates, and we intend to expand on them in our future work.

\bibliography{bibliography_conflict}

\end{spacing}

\clearpage

\section*{Appendix A: Additional Tables and Figures}\label{Appendix:TabFig}
\setcounter{equation}{0}
\setcounter{section}{0}
\setcounter{subsection}{0}
\setcounter{page}{1}
\renewcommand{\theequation}{\mbox{A\arabic{equation}}}
\renewcommand{\thesection}{\mbox{A\arabic{section}}}
\renewcommand{\thesubsection}{\mbox{A\arabic{subsection}}}
\renewcommand{\thetheorem}{\mbox{A\arabic{theorem}}}
\renewcommand{\theproposition}{\mbox{A\arabic{proposition}}}
\renewcommand{\thepage}{\mbox{A-\arabic{page}}}
\setcounter{table}{0}
\renewcommand{\thetable}{\mbox{A\arabic{table}}}
\setcounter{figure}{0}
\renewcommand{\thefigure}{\mbox{A\arabic{figure}}}

\begin{spacing}{1.0}
\begin{small}
%\clearpage
%\subsection{TABLES} 
\label{Tables}
%%%%%%%%%%%%%%%%%%% TABLES %%%%%%%%%%%%%%%%%%%%%%%%%%%%%%

\begin{table}[!htbp]\centering
\caption{Summary Statistics}\label{TABLE: Summary Statistics}
\vspace{-0.1cm}
\resizebox{0.93\textwidth}{!}{%
\begin{tabular}{lccccc}
\hline \hline

& Observations & Mean & SD & Min & Max \\ \hline 
%\vspace{-0.4cm}
& \multicolumn{5}{c}{\textit{Panel A: Interfirm Trade Data}} \\
%\vspace{0.4cm} \\
\cline{2-6}
\vspace{-0.4cm}
\input{Tables/Table_1A.tex}

\cline{2-6}
& \multicolumn{5}{c}{\textit{Panel B: Accounting Data}} \\
%\vspace{0.4cm} \\
\cline{2-6}
\vspace{-0.4cm}
\input{Tables/Table_1B.tex}

\hline \hline
\end{tabular}
}
%\\[0.2cm]
\vspace{0.8cm}
\begin{minipage}{0.93\textwidth}
\renewcommand{\baselinestretch}{0.7}
\footnotesize{\textit{Notes}: Interfirm trade intensity is measured by an indicator for any shipment from establishment (firm-rayon) $i$ to establishment $j$ at year-month $t$, the logarithm of the total number of shipments from establishment $i$ to establishment $j$ at year-month $t$, and the total weight of these shipments. To avoid missing values, the logarithms contain 1 inside. IHS stands for inverse hyperbolic sine transformation $L(X) = \log(X + \textnormal{sqrt}(X^2+1))$ as in \cite{mackinnon1990transforming}.\\
}
\end{minipage}
\end{table}

%\footnotesize{$\hat \beta^{(1)}$: (1 = Either partner is in the conflict area) $\times$ (1 = Post Crimea)}
%\footnotesize{$\hat \beta^{(2)}$: (1 = Either partner traded with conflict area) $\times$ (1 = Post Crimea)}

%Table AX
\begin{table}[!t]\centering
\caption{Propagation of the Conflict Shock: Firm-level Aggregation}\label{TABLE: Propagation Firm Level Aggregation}
\vspace{-0.1cm}
\resizebox{\textwidth}{!}{%
\begin{tabular}{lcccccc}
\hline
			&       (1)       &       (2)        &       (3)        &       (4)        &       (5)        &       (6)  \\
           &       Any        &       Log       &       Log Total        &       Any        &       Log       &       Log Total     \\  
           &       Shipment        &      Number of      &       Weight        &      Shipment       &     Number  of       &       Weight     \\  
            &              &      Shipments        &       Shipped        &            &    Shipments         &       Shipped     \\  
           \hline
           \\[-0.25cm]
\input{Tables/Table_1_Firm_Month}
\hline \hline 
\end{tabular}}
\\
\begin{minipage}{\textwidth}
\vspace{0.1cm}
\renewcommand{\baselinestretch}{0.7}
\footnotesize{\textit{Notes:} This table provides a version of the baseline results from equation~\eqref{eq: propagation} and Table \ref{TABLE: Reduced-Form Estimates of the Reduction in Trade} focusing on the \textit{firm-level} instead of the \textit{establishment-level} second-degree conflict treatment. The outcomes are: an indicator for any shipment from establishment (firm-rayon) $i$ to establishment $j$ at year-month $t$, the logarithm of the total number of shipments from establishment $i$ to establishment $j$ at year-month $t$, and the total weight of these shipments. To avoid missing values, the logarithms contain 1 inside. Standard errors in parentheses are clustered at the province-pair level. * p$<$0.05, ** p$<$0.01, *** p$<$0.001.\\
}
\end{minipage}
\end{table}

%Table AX
\begin{table}[!t]
\caption{Propagation of the Conflict Shock: \cite{borusyak2020non} Adjustment}\label{TABLE: Propagation Borusyak and Hull Adjustment}
\vspace{-0.1cm}
\resizebox{\textwidth}{!}{%
\begin{tabular}{lcccccc}
\hline
			&       (1)       &       (2)        &       (3)        &       (4)        &       (5)        &       (6)  \\
           &       Any        &       Log       &       Log Total        &       Any        &       Log       &       Log Total     \\  
           &       Shipment        &      Number of      &       Weight        &      Shipment       &     Number  of       &       Weight     \\  
            &              &      Shipments        &       Shipped        &            &    Shipments         &       Shipped     \\  
           \hline
           \\[-0.25cm]
\input{Tables/Table_1_Firm_Rayon_Month_BH}

\hline \hline 
\end{tabular}}
\\
\begin{minipage}{\textwidth}
\vspace{0.1cm}
\renewcommand{\baselinestretch}{0.7}
\footnotesize{\textit{Notes:} This table provides a version of the baseline results from equation~\eqref{eq: propagation} and Table \ref{TABLE: Reduced-Form Estimates of the Reduction in Trade} using an adjustment in the spirit of \cite{borusyak2020non} controlling for the number of supplier's and buyer's preconflict partners interacted with the postconflict indicator.  The outcomes are: an indicator for any shipment from establishment (firm-rayon) $i$ to establishment $j$ at year-month $t$, the logarithm of the total number of shipments from establishment $i$ to establishment $j$ at year-month $t$, and the total weight of these shipments. To avoid missing values, the logarithms contain 1 inside. Standard errors in parentheses are clustered at the province-pair level. * p$<$0.05, ** p$<$0.01, *** p$<$0.001.\\
}
\end{minipage}
\end{table}

%Table 2
\begin{table}[!htbp]\centering
\caption{Propagation of the Conflict Shock: Downstream and Upstream Effects}\label{TABLE: Propagation Downstream and Upstream}
\vspace{-0.1cm}
\resizebox{\textwidth}{!}{%
\begin{tabular}{lcccccc}
\hline
           &       (1)       &       (2)        &       (3)        &       (4)        &       (5)        &       (6)  \\
           &       Any        &       Log       &       Log Total        &       Any        &       Log       &       Log Total     \\  
           &       Shipment        &      Number of      &       Weight        &      Shipment       &     Number  of       &       Weight     \\  
            &              &      Shipments        &       Shipped        &            &    Shipments         &       Shipped     \\  
           \hline
           \\[-0.25cm]
\input{Tables/Table_2_Firm_Rayon_Month_UD}
\hline \hline 
\end{tabular}

}
\\
\begin{minipage}{\textwidth}
\vspace{0.2cm}
\renewcommand{\baselinestretch}{0.7}
\footnotesize{\textit{Notes:} This table presents the estimates of equation~\eqref{eq: propagation separated} studying whether interfirm trade declined after the start of the Russia-Ukraine conflict depending on whether one of the partners in an exchange located in or have at any point traded with the conflict area. In contrast to equation~\eqref{eq: propagation}, this specification separates between upstream and downstream shock propagation. The outcomes are: an indicator for any shipment from establishment (firm-rayon) $i$ to establishment $j$ at year-month $t$, the logarithm of the total number of shipments from establishment $i$ to establishment $j$ at year-month $t$, and the total weight of these shipments. To avoid missing values, the logarithms contain 1 inside. All specifications exclude trade links in which both partners are located in the conflict areas. Standard errors in parentheses are clustered at the province-pair level. * p$<$0.05, ** p$<$0.01, *** p$<$0.001.\\
}
\end{minipage}
\end{table}

\begin{table}[!t]\centering
\caption{Propagation of the Conflict Shock: Controlling For Firms' Distance to Conflict Areas}\label{TABLE: Propagation Controlling For Distance}
\vspace{-0.1cm}
\resizebox{\textwidth}{!}{%
\begin{tabular}{lcccccc}
\hline
			&       (1)       &       (2)        &       (3)        &       (4)        &       (5)        &       (6)  \\
           &       Any        &       Log       &       Log Total        &       Any        &       Log       &       Log Total     \\  
           &       Shipment        &      Number of      &       Weight        &      Shipment       &     Number  of       &       Weight     \\  
            &              &      Shipments        &       Shipped        &            &    Shipments         &       Shipped     \\  
           \hline
           \\[-0.25cm]
\input{Tables/Table_1_Firm_Rayon_Month_Dist}
\hline \hline 
\end{tabular}}
\\
\begin{minipage}{\textwidth}
\vspace{0.1cm}
\renewcommand{\baselinestretch}{0.7}
\footnotesize{\textit{Notes:} This table provides a version of the baseline results from equation~\eqref{eq: propagation} and Table \ref{TABLE: Reduced-Form Estimates of the Reduction in Trade} but flexibly controlling for the sender's and receiver's spherical distance to the conflict areas. The outcomes are: an indicator for any shipment from establishment (firm-rayon) $i$ to establishment $j$ at year-month $t$, the logarithm of the total number of shipments from establishment $i$ to establishment $j$ at year-month $t$, and the total weight of these shipments. To avoid missing values, the logarithms contain 1 inside. Standard errors in parentheses are clustered at the province-pair level. * p$<$0.05, ** p$<$0.01, *** p$<$0.001.\\
}
\end{minipage}
\end{table}

\begin{table}[!t]\centering
\caption{Propagation of the Conflict Shock: Controlling For Post-Province FE}\label{TABLE: Propagation Controlling For Province FE}
\vspace{-0.1cm}
\resizebox{\textwidth}{!}{%
\begin{tabular}{lcccccc}
\hline
			&       (1)       &       (2)        &       (3)        &       (4)        &       (5)        &       (6)  \\
           &       Any        &       Log       &       Log Total        &       Any        &       Log       &       Log Total     \\  
           &       Shipment        &      Number of      &       Weight        &      Shipment       &     Number  of       &       Weight     \\  
            &              &      Shipments        &       Shipped        &            &    Shipments         &       Shipped     \\  
           \hline
           \\[-0.25cm]
\input{Tables/Table_1_Firm_Rayon_Month_PostPrFE.tex}
\hline \hline 
\end{tabular}}
\\
\begin{minipage}{\textwidth}
\vspace{0.1cm}
\renewcommand{\baselinestretch}{0.7}
\footnotesize{\textit{Notes:} This table provides a version of the baseline results from equation~\eqref{eq: propagation} and Table \ref{TABLE: Reduced-Form Estimates of the Reduction in Trade} but controlling for the post-Crimea indicator interacted with the province indicators. The outcomes are: an indicator for any shipment from establishment (firm-rayon) $i$ to establishment $j$ at year-month $t$, the logarithm of the total number of shipments from establishment $i$ to establishment $j$ at year-month $t$, and the total weight of these shipments. To avoid missing values, the logarithms contain 1 inside. Standard errors in parentheses are clustered at the province-pair level. * p$<$0.05, ** p$<$0.01, *** p$<$0.001.\\
}
\end{minipage}
\end{table}

\begin{table}[!htbp]\centering
\caption{Effect of Conflict on Firm Performance Through Change in Network Structure}\label{TABLE: Centrality}
\vspace{-0.1cm}
\resizebox{\textwidth}{!}{%
\begin{tabular}{lcccccc}
\hline
           &      \multicolumn{3}{c}{Logarithm Sales}       &      \multicolumn{3}{c}{IHS Profits}     \\  
           &       (1)       &       (2)        &       (3)        &       (4)        &       (5)        &       (6)  \\
           \hline
           \\[-0.35cm]
%                    &       %est1        &       %est2        &       %est3        &       %est4        &       %est5        &       %est6        \\[-0.1cm]
Post Crimea $\times$ Change in Eigenvector Centrality &       0.145$^{***}$&                    &                    &       0.679$^{**}$ &                    &                    \\[-0.1cm]
	  &     (0.039)        &                    &                    &     (0.262)        &                    &                    \\[0.1cm]
Post Crimea $\times$ Change in Betweenness Centrality &                    &       0.118$^{***}$&                    &                    &       0.684$^{***}$&                    \\[-0.1cm]
                   &                    &     (0.030)        &                    &                    &     (0.195)        &                    \\[0.1cm]
Post Crimea $\times$ Change in Degree Centrality  &                    &                    &       0.089$^{**}$ &                    &                    &       0.382$^{*}$  \\[-0.1cm]
                   &                    &                    &     (0.028)        &                    &                    &     (0.192)        \\[0.1cm]\addlinespace[0.15cm]
Firm FE        &{\checkmark}        &{\checkmark}        &{\checkmark}        &{\checkmark}        &{\checkmark}        &{\checkmark}        \\[-0.15cm]
Year FE       &{\checkmark}        &{\checkmark}        &{\checkmark}        &{\checkmark}        &{\checkmark}        &{\checkmark}        \\[-0.15cm]\addlinespace[0.15cm]

Mean                &      17.062        &      17.062        &      17.062        &       6.941        &       6.941        &       6.941        \\[-0.15cm]
SD                  &       2.444        &       2.444        &       2.444        &      13.036        &      13.036        &      13.036        \\[-0.15cm]
R$^2$               &        0.81        &        0.81        &        0.81        &        0.47        &        0.47        &        0.47        \\[-0.15cm]
Observations        &      30,209        &      30,209        &      30,209        &      29,495        &      29,495        &      29,495        \\[-0.15cm]
Number of Firms     &       3,997        &       3,997        &       3,997        &       3,970        &       3,970        &       3,970        \\[-0.15cm]
\addlinespace[0.15cm] 
\hline \hline 
\end{tabular}

}
\\
\begin{minipage}{\textwidth}
\vspace{0.2cm}
\renewcommand{\baselinestretch}{0.7}
\footnotesize{\textit{Notes:} This table presents the estimates of equation~\eqref{eq: centrality} studying whether firms that for exogenous reasons experienced a change in their production-network centrality changed their relative performance after the start of the conflict. All measures of centrality change are standardized to have mean zero and standard deviation of one. The sample is restricted to firms outside of the conflict areas (i.e., DPR, LPR, and Crimea). The dependent variable in Columns (1)--(3) is the logarithm of sales calculated as $L(X) = log(1+X)$. The dependent variable in Columns (4)--(6) is the inverse hyperbolic sine of gross profit calculated as $L(X) = log(X + \sqrt{X^2+1})$ following \cite{mackinnon1990transforming}. Standard errors in parentheses are clustered at the firm level. * p$<$0.05, ** p$<$0.01, *** p$<$0.001. \\
}
\end{minipage}
\end{table}

%
%\begin{table}[!htbp]
%\caption{Counterfactuals---Changes}\label{table:counterfactuals_2}
%
%\resizebox{0.85\textwidth}{!}{\begin{tabular}{lcccccc}
%\hline
%\vspace{-0.2cm}
% & (1) & (2) & (3) & (4) & (5) & (6) \\ \vspace{.15cm}
% & $\Delta Q_2 \%$ & $\Delta Mean \%$ & $N$ & $\Delta Q_2 \%$ & $\Delta Mean \%$ & $N$ \\ \hline \vspace{.15cm} 
%
%
%\vspace{-.15cm}  & \multicolumn{3}{l}{\textit{Panel A: $|\Delta|<110\%$}} & \multicolumn{3}{l}{\textit{Panel B: $|\Delta|<110\%$, $\Delta \neq 0$}} \\
%
%$\frac{\bold{r}^{2014}-\bold{r}^{2013}}{\bold{r}^{2013}}$ &      -0.1 &      -9.6 &     4,117 &      -0.2 &      -9.6 &     4,114 \\
%$\frac{\bold{r}^{EN}-\bold{r}^{2013}}{\bold{r}^{2013}}$ &       0.0 &      -7.0 &     3,880 &      -2.9 &     -16.4 &     1,671 \\ 
%$\frac{\bold{r}^{FD}-\bold{r}^{2013}}{\bold{r}^{2013}}$  &      -3.0 &     -10.3 &     2,941 &      -3.1 &     -10.3 &     2,939 \\ 
%$\frac{\bold{r}^{NA}-\bold{r}^{2013}}{\bold{r}^{2013}}$ &       0.0 &      -2.3 &     3,393 &      -0.5 &      -4.6 &     1,666 \\ 
%
%\hline
%\end{tabular}
%}
%
%\begin{minipage}{\textwidth}
%\vspace{0.2cm}
%\renewcommand{\baselinestretch}{0.7}
%\footnotesize{\textit{Notes:} \\
%}
%\end{minipage}
%
%\end{table}

%
%%%%%%%%%%%%%%%%%%% FIGURES %%%%%%%%%%%%%%%%%%%%%%%%%%%%%%
%\clearpage
%\section*{FIGURES}

%%Figure 2
%\begin{figure}[!htbp]
%\begin{center}
%\begin{minipage}{\textwidth}
%    \begin{center}
%        \includegraphics[width=0.91\textwidth]{1_EventStudy_log_weight.pdf}\\
%    \end{center}
%    \vspace{-0.2cm}
%\caption{Event Study Graph, Pooled}\label{FIG: Event Study 1} %
%\end{minipage}
%\end{center}
%\end{figure}

\clearpage

\begin{table}[!]
\caption{Counterfactuals---Other Features of Distributions, Baseline Sample}\label{TABLE:counter_expanded}

\resizebox{0.95\textwidth}{!}{
\begin{tabular}{lccccc}
\hline
 & (1) & (2) & (3) & (4) & (5) \\

& $\bold{r}(\bold{\Omega^{2013}}, \bm{\xi^{2013}})$ & $\bold{r}(\bold{\tilde \Omega^{2013}},\bm{\xi^{2013}})$ & $\bold{r}(\bold{\Omega^{2014}},\bm{\xi^{2013}})$
 &  $\bold{r}(\bold{\Omega^{2013}},\bm{\xi^{2014}})$ & $\bold{r}(\bold{\Omega^{2014}},\bm{\xi^{2014}})$   \\

\hline

& \makecell{Baseline \\ \ } & \makecell{Network \\ Destruction} & \makecell{Network \\ Adjustment}  & \makecell{Outside \\ Demand} & \makecell{Total \\ Change}    \\ \hline

% Median $\bold{r}$, Mln. Hr. & 17.9 & 9.5 & 16.1 & 14.8 & 13.0  \\

% \medskip

% Relative to 2013 & 0.0\% & -46.8\% & -9.7\% & -17.3\% & -27.2\% \\ 

% 25th \%-tile $\bold{r}$, Mln. Hr. & 4.2 & 0.3 & 1.6 & 1.0 & 1.6 \\ 

% \medskip

% Relative to 2013 & 0.0\% & -92.9\% & -62.3\% & -76.0\% & -62.0\% \\ 

% 75th \%-tile $\bold{r}$, Mln. Hr. & 73.7 & 52.1 & 84.1 & 76.0 & 56.9 \\ 

% \medskip

%  Relative to 2013 & 0.0\% & -29.3\% & 14.0\% & 3.1\% & -22.8\% \\ 
 
% Mean $\bold{r}$, Mln. Hr. & 240.2 & 95.4 & 213.8 & 204.6 & 198.4 \\ 

%   Relative to 2013 & 0.0\% & -60.3\% & -11.0\% & -14.8\% & -17.4\%  \\
  
% \hline

% $N$ & 4,463 & 4,463 & 4,463 & 4,463 & 4,463 \\ 

Median $\bold{r}$ & 2.25 & 1.20 & 2.03 & 1.86 & 1.64 \\

\medskip

Relative to 2013 & 0.0\% & -46.8\% & -9.7\% & -17.3\% & -27.2\%\\

25th \%-tile $\bold{r}$ & 0.54 & 0.04 & 0.20 & 0.13 & 0.20\\

\medskip

Relative to 2013 &  0.0\% & -92.9\% & -62.3\% & -76.0\% & -62.0\%\\

75th \%-tile $\bold{r}$ & 9.29 & 6.57 & 10.59 & 9.58 & 7.17 \\

\medskip

 Relative to 2013 & 0.0\% & -29.3\% & 14.0\% & 3.1\% & -22.8\%\\
 
Mean $\bold{r}$ & 30.27 & 12.02 & 26.94 & 25.78 & 24.99\\

  Relative to 2013 & 0.0\% & -60.3\% & -11.0\% & -14.8\% & -17.4\%\\
  
\hline
  
$N$ & 4,463 & 4,463 & 4,463 & 4,463 & 4,463 \\ 

\hline

\end{tabular}
}

\begin{minipage}{\textwidth}
\vspace{0.2cm}
\renewcommand{\baselinestretch}{0.7}
\footnotesize{\textit{Notes:} The table presents the extended version of the counterfactual estimates in Table~\ref{table:counterfactuals}. Specifically, we obtain the vector of revenues $\bm{r}$ for a given input-output matrix $\bold{\Omega}$ and outside demand $\bm{\xi}$ such that    $\bm{r}(\bm{\xi},\bold{\Omega}):=[\bold{I}-(1-\alpha)\bold{\Omega}]^{-1}\bm{\xi}$ and present four summary statistics of the resulting distributions: median, 25th percentile, 75th percentile, and the mean. Columns vary each of the arguments. Column (1) displays the actual revenue statistic in 2013. Column (2) uses the modified 2013 input-output matrix where trade with firms from the conflict areas is replaced with zeros, $\bold{\tilde \Omega^{2013}}$. Column (3) fixes the outside demand at the 2013 level and uses the actual 2014 input-output matrix, $\bold{\Omega^{2014}}$. Column (4) uses the 2014 outside demand and the 2013 input-output matrix, $\bold{\Omega^{2013}}$. Column (5) displays the actual revenue statistic in 2014. The sample is restricted to firms in nonconflict areas. All quantities are expressed in 2010 \$US.
}
\end{minipage}
\end{table}

\clearpage

\begin{table}[!]
\caption{Counterfactuals---Other Features of Distributions, Sample of All Firms}\label{TABLE:counter_sample}

\resizebox{0.95\textwidth}{!}{
\begin{tabular}{lccccc}
\hline
 & (1) & (2) & (3) & (4) & (5) \\

& $\bold{r}(\bold{\Omega^{2013}}, \bm{\xi^{2013}})$ & $\bold{r}(\bold{\tilde \Omega^{2013}},\bm{\xi^{2013}})$ & $\bold{r}(\bold{\Omega^{2014}},\bm{\xi^{2013}})$
 &  $\bold{r}(\bold{\Omega^{2013}},\bm{\xi^{2014}})$ & $\bold{r}(\bold{\Omega^{2014}},\bm{\xi^{2014}})$   \\

\hline

& \makecell{Baseline \\ \ } & \makecell{Network \\ Destruction} & \makecell{Network \\ Adjustment}  & \makecell{Outside \\ Demand} & \makecell{Total \\ Change}    \\ \hline

% Median $\bold{r}$, Mln. Hr. & 14.7 & 7.4 & 13.5 & 12.4 & 10.8 \\ 

% \medskip

% Relative to 2013 & 0.0\% & -50.1\% & -8.6\% & -16.1\% & -26.5\% \\

% 25th \%-tile $\bold{r}$, Mln. Hr. & 2.7 & 0.0 & 0.7 & 0.5 & 0.9 \\  

% \medskip

% Relative to 2013 & 0.0\% & -100.0\% & -74.9\% & -82.1\% & -65.7\% \\  

% 75th \%-tile $\bold{r}$, Mln. Hr. & 64.8 & 45.1 & 77.7 & 69.4 & 52.1 \\ 

% \medskip

% Relative to 2013 & 0.0\% & -30.4\% & 20.0\% & 7.2\% & -19.5\% \\  

% Mean $\bold{r}$, Mln. Hr. & 222.7 & 87.6 & 204.5 & 191.8 & 186.0 \\ 

% Relative to 2013 & 0.0\% & -60.7\% & -8.1\% & -13.8\% & -16.5\% \\ 

\hline

Median $\bold{r}$ & 1.85 & 0.93 & 1.70 & 1.56 & 1.36 \\ 

\medskip

Relative to 2013 & 0.0\% & -50.1\% & -8.6\% & -16.1\% & -26.5\% \\

25th \%-tile $\bold{r}$ & 0.34 & 0.0 & 0.09 & 0.06 & 0.11 \\  

\medskip

Relative to 2013 & 0.0\% & -100.0\% & -74.9\% & -82.1\% & -65.7\% \\  

75th \%-tile $\bold{r}$ & 8.17 & 5.69 & 9.80 & 8.75 & 6.57 \\ 

\medskip

Relative to 2013 & 0.0\% & -30.4\% & 20.0\% & 7.2\% & -19.5\% \\  

Mean $\bold{r}$ & 28.1 & 11.05 & 25.79 & 24.19 & 23.46 \\ 

Relative to 2013 & 0.0\% & -60.7\% & -8.1\% & -13.8\% & -16.5\% \\ 

\hline

$N$ & 4,815 & 4,815 & 4,815 & 4,815 & 4,815 \\
  
\hline

\end{tabular}
}

\begin{minipage}{\textwidth}
\vspace{0.2cm}
\renewcommand{\baselinestretch}{0.7}
\footnotesize{\textit{Notes:} The table presents a modified version of the counterfactual estimates in Table~\ref{TABLE:counter_expanded} calculated for an expanded sample of firms. Specifically, in these estimates, the sample is all firms in nonconflict areas that appeared in the data at least once, thus accounting for firm entry. We obtain the vector of revenues $\bm{r}$ for a given input-output matrix $\bold{\Omega}$ and outside demand $\bm{\xi}$ such that    $\bm{r}(\bm{\xi},\bold{\Omega}):=[\bold{I}-(1-\alpha)\bold{\Omega}]^{-1}\bm{\xi}$ and present four summary statistics of the resulting distributions: median, 25th percentile, 75th percentile, and the mean. Columns vary each of the arguments. Column (1) displays the actual revenue statistic in 2013. Column (2) uses the modified 2013 input-output matrix where links with firms from the conflict areas are replaced by zeros, $\bold{\tilde \Omega^{2013}}$. Column (3) fixes the outside demand at the 2013 level and uses the actual 2014 input-output matrix, $\bold{\Omega^{2014}}$. Column~(4) uses the 2014 outside demand and the 2013 input-output matrix, $\bold{\Omega^{2013}}$. Column~(5) displays the actual revenue statistic in 2014. All quantities are expressed in 2010 \$ US.
}
\end{minipage}
\end{table}

\clearpage
\newpage
\begin{table}[!]
\caption{Province- and District-Level Counterfactuals}\label{table:regions}

\resizebox{0.75\textwidth}{!}{\begin{tabular}{lcccccc}
\hline
 & (1) & (2) & (3) & (4) & (5) & (6) \\

& Raw & Net.Adj. & Raw. & Raw & Net.Adj. & Raw. \\
& 2013 & 2014 & 2014 & 2013 & 2014 & 2014 \\ \hline

 & \multicolumn{3}{c}{\textit{Panel A: Province Level}} & \multicolumn{3}{c}{\textit{Panel B: District Level}} \\
Total Revenue & 431,806 & 525,712 & 382,843 & 443,296 & 516,716 & 399,908 \\
Relative to 2013 & 0.0\% & 21.7\% & -11.3\% & 0.0\% & 16.6\% & -9.8\%  \\
$N$ & 23 & 23 & 23 & 402 &	402 & 402 \\

%  & \multicolumn{3}{c}{\textit{Panel A: Province Level}} & \multicolumn{3}{c}{\textit{Panel B: District Level}} \\
% Total Revenue & 431,806.1 & 525,712.3 & 382,842.6 & 443,296.2 & 516,716.2 & 399907.7 \\
% Relative to 2013 & 0.0\% & 21.7\% & -11.3\% & 0.0\% & 16.6\% & -9.8\%   \\
% $N$ & 23 & 23 & 23 & 402 & 402 & 402 \\

\hline

\end{tabular}
}

\begin{minipage}{0.76\textwidth}
\vspace{0.2cm}
\renewcommand{\baselinestretch}{0.7}
\footnotesize{\textit{Notes:} The table presents the aggregate results for Ukrainian provinces (oblasts) and districts (rayons). We use 1,334,052 firm-year observations from 2013 through 2016 to calculate province- and district-level total revenues. Columns~(1) and~(4) display the actual total revenues in 2013 at the province and district levels, respectively. Columns~(2) and~(5) obtain the counterfactual vector of locality revenues $\bm{r}$ for a given input-output matrix $\bold{\Omega}$ and outside demand $\bm{\xi}$ such that    $\bm{r}(\bm{\xi},\bold{\Omega}):=[\bold{I}-(1-\alpha)\bold{\Omega}]^{-1}\bm{\xi}$, fixing outside demand at the 2013 level and using the actual 2014 railway trade matrix between geographic units, $\bold{\Omega^{2014}}$. Columns~(3) and~(6) display the actual total revenues in 2014 at the province and district levels, respectively. The sample is restricted to localities not exposed to violence directly. All quantities are expressed in 2010 \$US.
}
\vspace{-0.3cm}
\end{minipage}
\end{table}

%Figure 7
\clearpage
\newpage
\begin{figure}[!htbp]
\begin{center}
\begin{minipage}{0.8\textwidth}
    \begin{center}
        \includegraphics[width=0.95\textwidth]{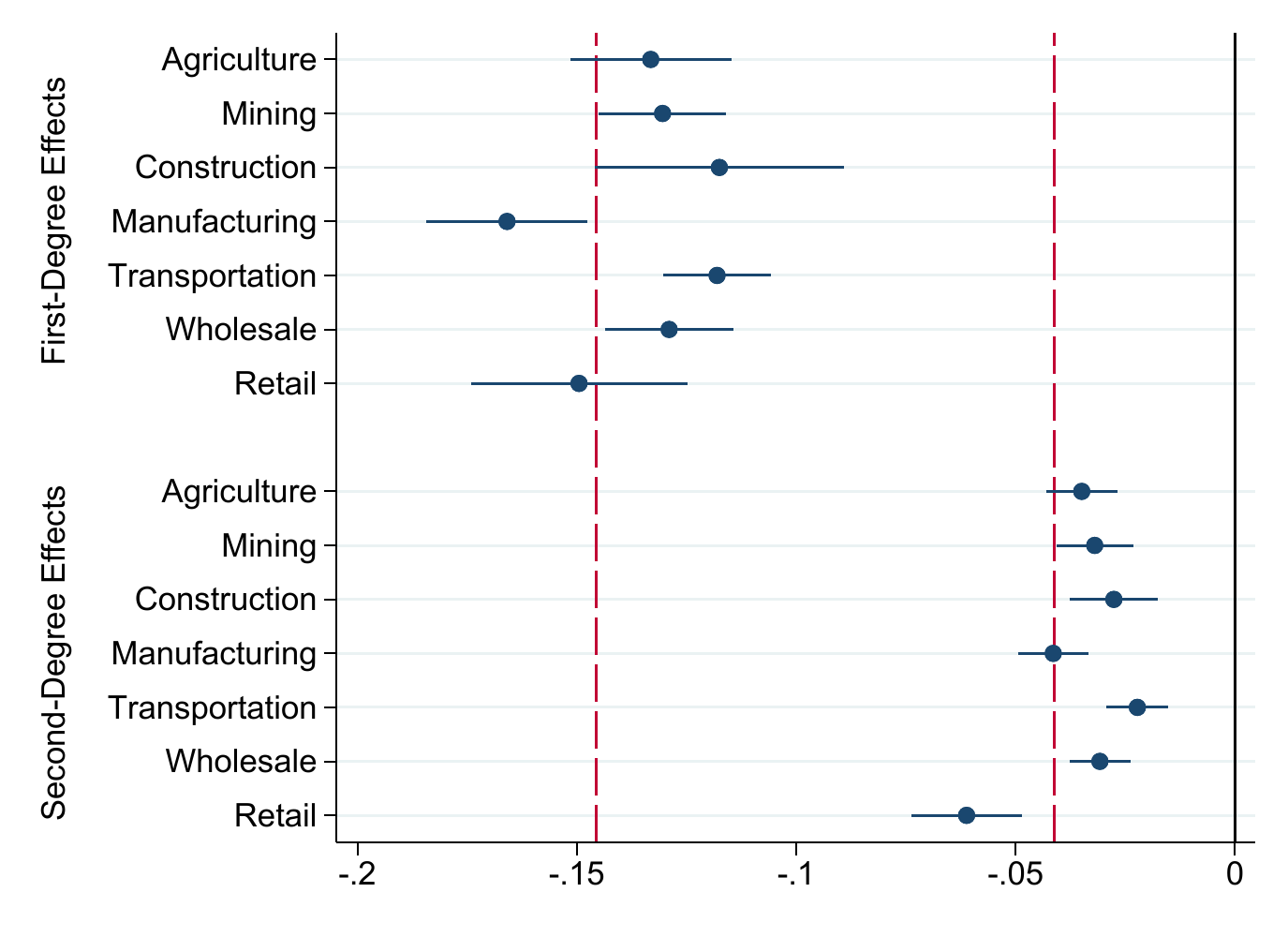}\\
    \end{center}
    \vspace{-0.2cm}
\caption{Heterogeneity of the Effects of Conflict on Trade by Firms' Industry}\label{FIG: Propagation Heterogeneity by Industry} %
\vspace{-0.2cm}
\renewcommand{\baselinestretch}{0.7}
\footnotesize{\textit{Notes}: The figure presents the heterogeneity estimates of the propagation estimates in Table~\ref{TABLE: Reduced-Form Estimates of the Reduction in Trade} by firms' industry identified through their SIC classification. The outcome is the monthly frequency of trade. The figure is constructed as follows. First, we estimate the coefficients on the interactions between the first- and second-degree-conflict-partner indicators and the industry indicators for the seven SIC industry categories composing more than 3\% of the sample (i.e., whether at least one of the two trade partners in a link works in one of these industries). Next, we linearly add the baseline and the heterogeneity postconflict first-degree and second-degree coefficients and depict them with 95\% confidence bands. All regressions include establishment-pair-direction and year-month fixed effects. All other specification details in the notes to Table~\ref{TABLE: Reduced-Form Estimates of the Reduction in Trade} apply.}
\end{minipage}
\end{center}
\end{figure}

%Figure 7
\clearpage
\newpage
\begin{figure}[!htbp]
\begin{center}
\begin{minipage}{0.7\textwidth}
    \begin{center}
        \includegraphics[width=\textwidth]{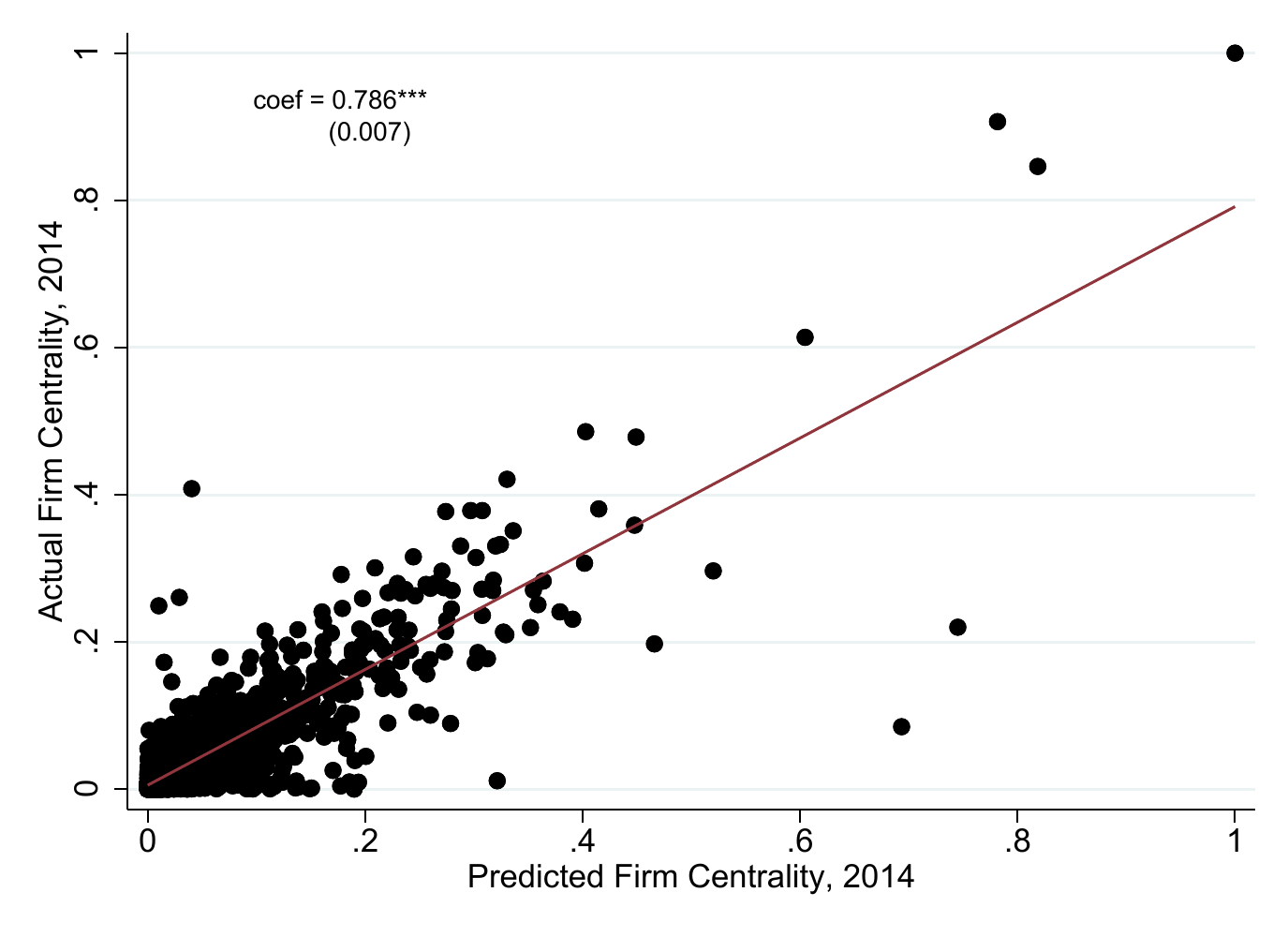}\\
    \end{center}
    \vspace{-0.2cm}
\caption{Relationship between Actual and Predicted Firm Centrality}\label{FIG: First Stage} %
\vspace{-0.2cm}
\renewcommand{\baselinestretch}{0.7}
\footnotesize{\textit{Notes}: This figure plots firm centrality in the 2014 production network with respect to predicted firm centrality estimated based on their network position in the modified 2013 production network without Crimea and the conflict areas in Donbas.}
\end{minipage}
\end{center}
\end{figure}

%Figure 5
\captionsetup{justification=centering}
\begin{figure}[!htbp]
\begin{center}
\begin{minipage}{0.65\textwidth}
    \begin{center}
        \includegraphics[width=0.9\textwidth]{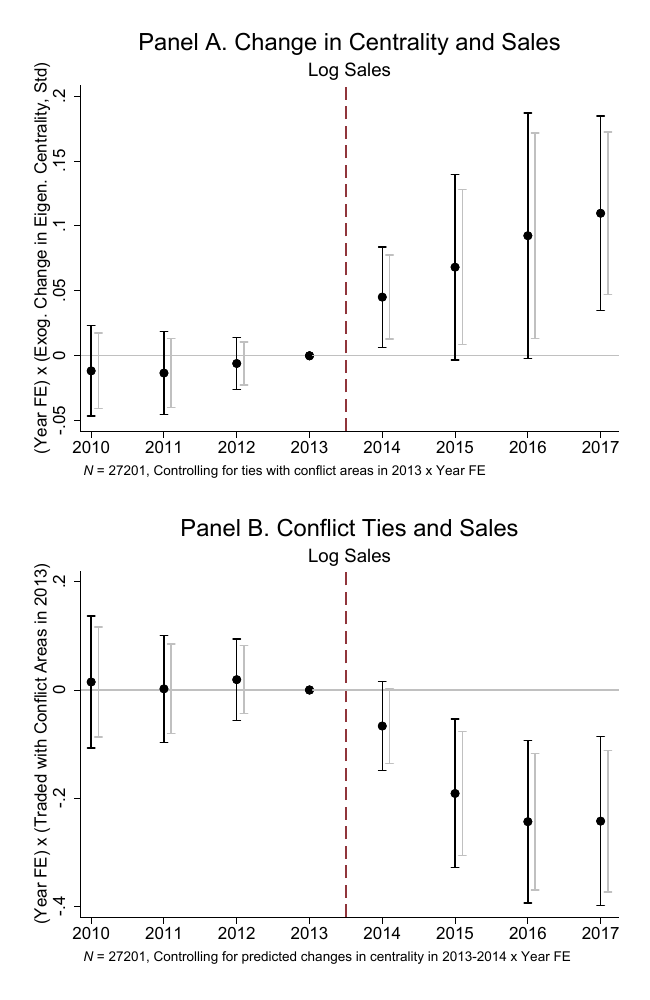}\\
    \end{center}
    \vspace{-0.2cm}
\caption{Conflict-Induced Change in Network Centrality and Sales: \\ Controlling for Ties to Conflict}\label{FIG: Sales, Network Centrality and Conflict} %
%\vspace{-0.2cm}
\renewcommand{\baselinestretch}{0.7}
\footnotesize{\textit{Notes}: This figure displays the results of estimating equation \eqref{eq: joint} and explores whether the baseline estimates for exogenous changes in firm centrality presented in \Cref{FIG: Sales and Centrality 1} and \Cref{TABLE: Centrality} are driven by the firm's prior trade ties with the conflict areas. Panel~A displays the results for the exogenous change in eigenvector centrality between 2013 and 2014 as the interaction variable, standardized to have zero mean and standard deviation of one. Panel~B displays the results for any trade ties with conflict areas in 2013 as the interaction variable. The outcome variable is the logarithm of sales. Black bars represent 95\% confidence intervals, gray bars represent 90\% confidence intervals. Standard errors are clustered at the firm level.}
\end{minipage}
\end{center}
\end{figure}

\clearpage
\newpage

\begin{figure}[!h]
\begin{center}
\begin{minipage}{0.52\textwidth}
    \begin{center}
        \includegraphics[width=\textwidth]{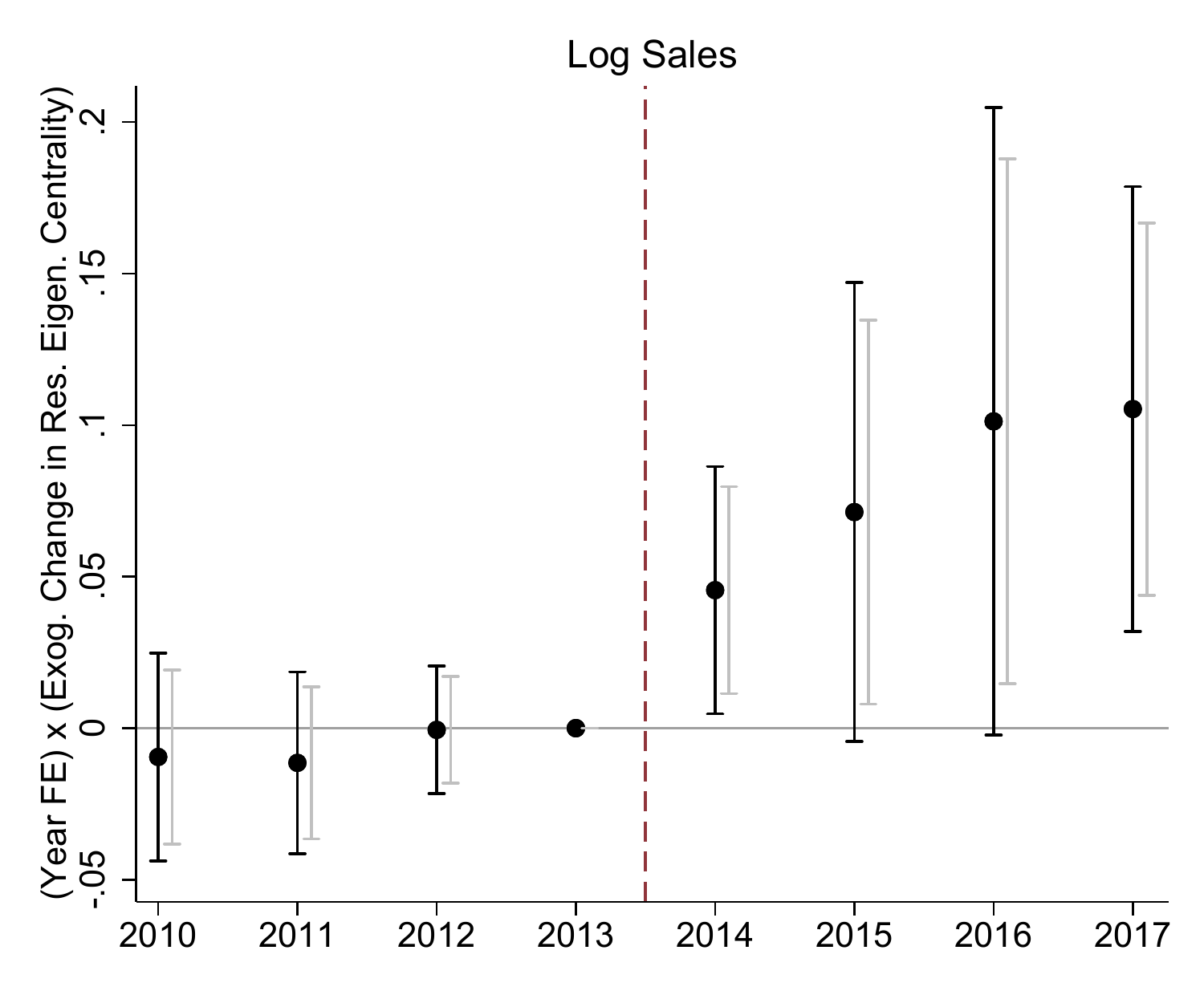}\\
    \end{center}
    \vspace{-0.2cm}
\caption{Conflict-Induced Change in Network Centrality and Sales: \\ Residualizing Firm Characteristics}\label{FIG: Residualizing Change in Centrality} %
\vspace{-0.2cm}
\end{minipage}
\begin{minipage}{0.99\textwidth}
\renewcommand{\baselinestretch}{0.7}
\footnotesize{\textit{Notes}: This figure explores whether the baseline estimates for changes in firm centrality presented in \Cref{FIG: Sales and Centrality 1} and \Cref{TABLE: Centrality} are driven by firm characteristics. We calculate the residual change in centrality as the residual after regressing the change in centrality without adjustment on: an indicator whether a firm traded with someone in the conflict areas, the share of its transactions with the conflict areas, the share of its total sales with the conflict areas, the total weight of a firm's partners' trade with the conflict areas (i.e., through second-degree connections), as well as firm's log sales and log profits in 2013. The obtained residuals are then used as a measure of exogenous change in centrality in equation~\eqref{eq: centrality}. The outcome variable is the logarithm of sales. Black bars represent 95\% confidence intervals, gray bars represent 90\% confidence intervals. Standard errors are clustered at the firm level.}
\end{minipage}
\end{center}
\end{figure}

\begin{figure}[!h]
\begin{center}
\begin{minipage}{\textwidth}
    \begin{center}
    \vspace{-0.4cm}
        \includegraphics[width=\textwidth]{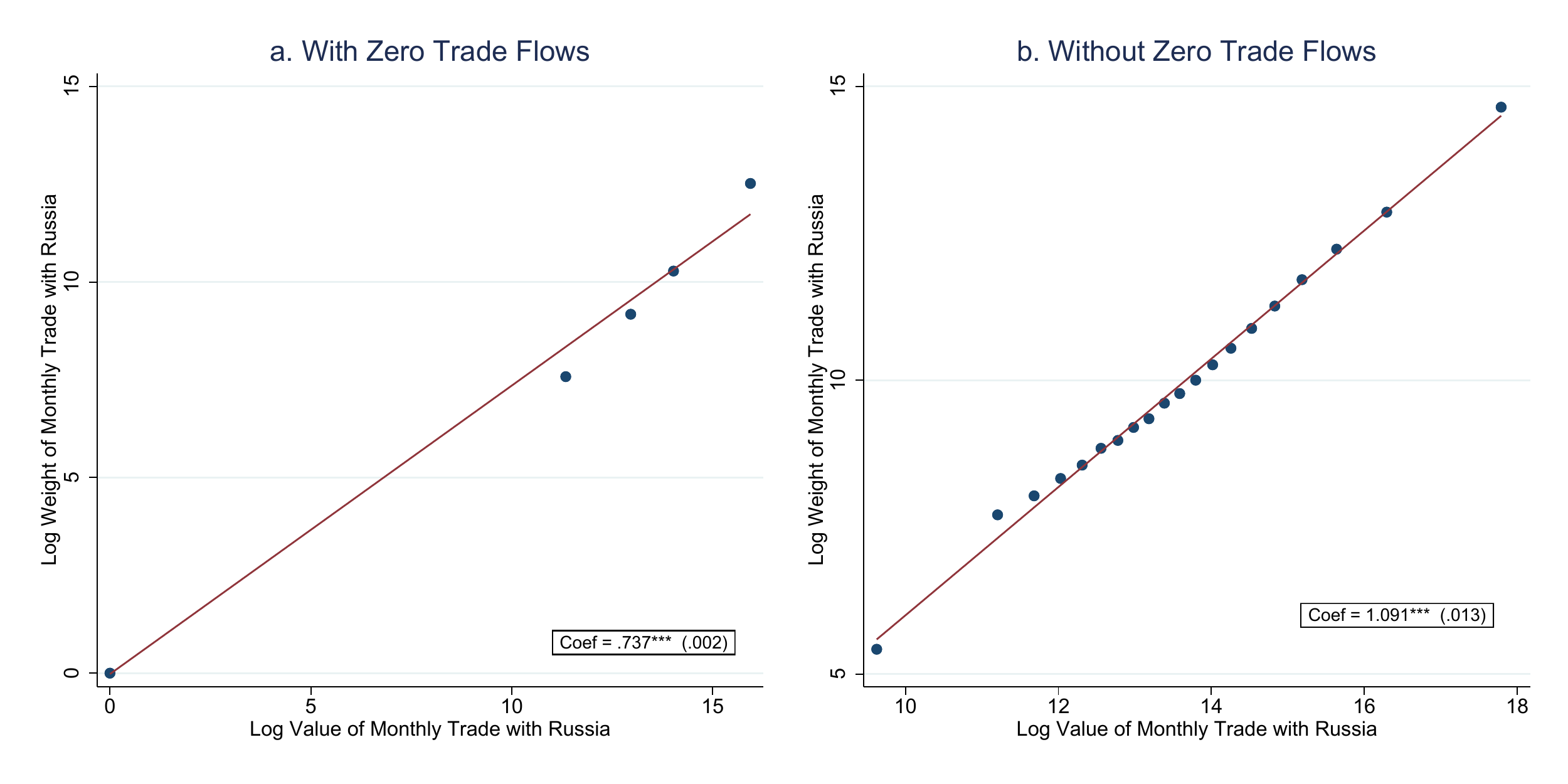}\\
    \end{center}
    \vspace{-0.2cm}
\caption{Correlation Between Total Weight and Value of Monthly Trade Flows Between Ukrainian Firms and Russia}\label{FIG: Weight Value Correlation} %
\vspace{-0.2cm}
\end{minipage}
\begin{minipage}{0.99\textwidth}
\renewcommand{\baselinestretch}{0.7}
\footnotesize{\textit{Notes}: This figure displays the binned scatter plots of the total weight and value of monthly trade flows between Ukrainian firms and Russia. Panel A includes the instances of zero monthly trade and Panel B excludes such instances. The figure relies on Ukrainian customs data from \cite{korovkin2019trading}. The time period is January 2013 through December 2016. Standard errors are clustered at the firm level. * p$<$0.05, ** p$<$0.01, *** p$<$0.001.}
\end{minipage}
\end{center}
\end{figure}

\begin{figure}[!h]
\begin{center}
\begin{minipage}{0.55\textwidth}
    \begin{center}
    \vspace{-0.4cm}
        \includegraphics[width=\textwidth]{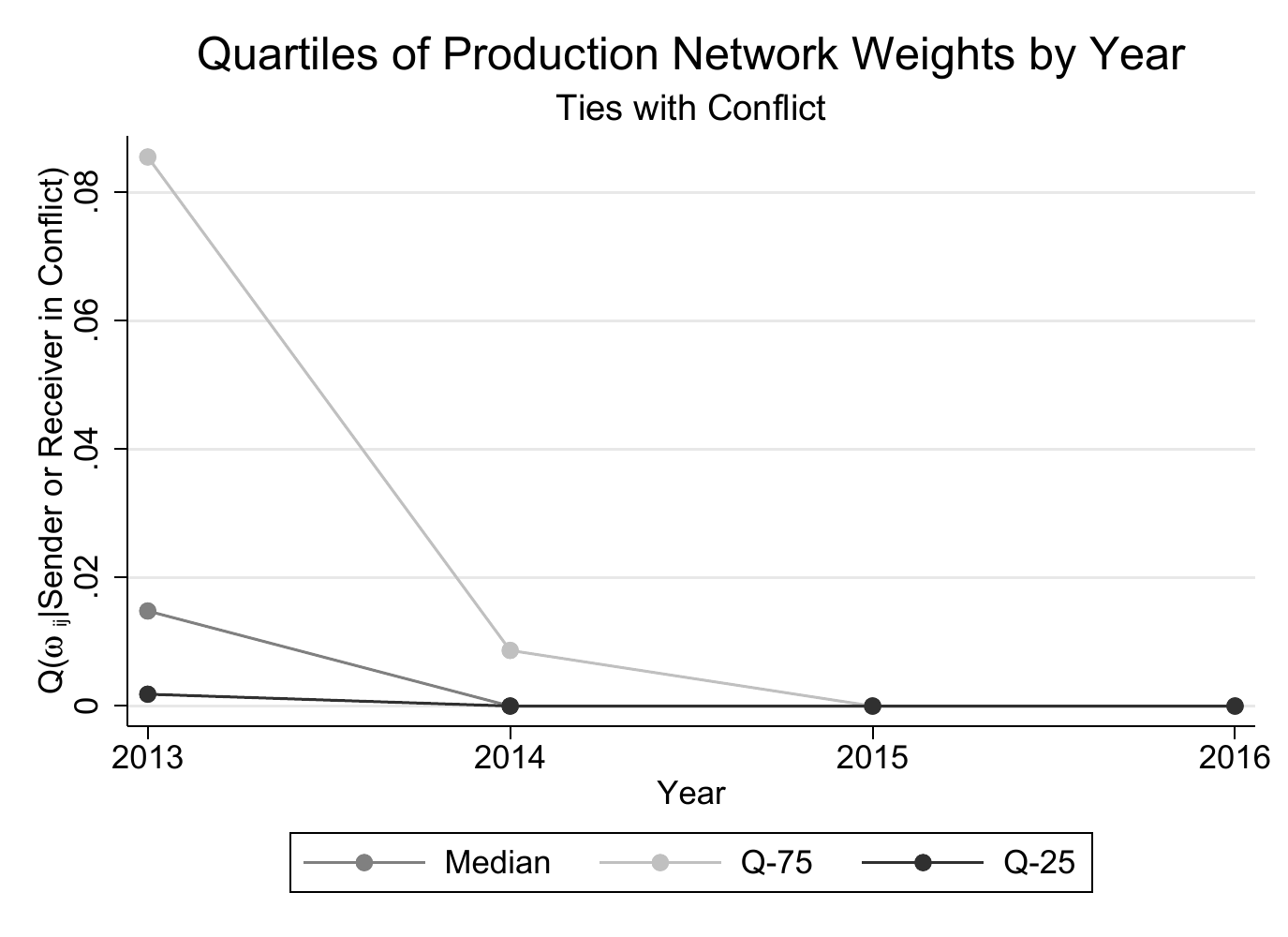}\\
    \end{center}
    \vspace{-0.2cm}
\caption{Production Network Weights for Trade with Conflict Areas}\label{FIG: Conflict Production Network Weights} %
\vspace{-0.2cm}
\end{minipage}
\begin{minipage}{0.99\textwidth}
\renewcommand{\baselinestretch}{0.7}
\footnotesize{\textit{Notes}: This figure displays the quartiles of the distribution of production network weights (elements) corresponding to trade with DPR, LPR, and Crimea. Q-25 refers to the first quartile or the 25th percentile and Q-75 refers to the third quartile or the 75th percentile of the element. distribution.}
\end{minipage}
\end{center}
\end{figure}

\clearpage
\newpage

\begin{figure}[!t]
\begin{center}
\begin{minipage}{0.95\textwidth}
    \begin{center}
        \vspace*{0.1cm} 
        \includegraphics[width=1.0\textwidth]{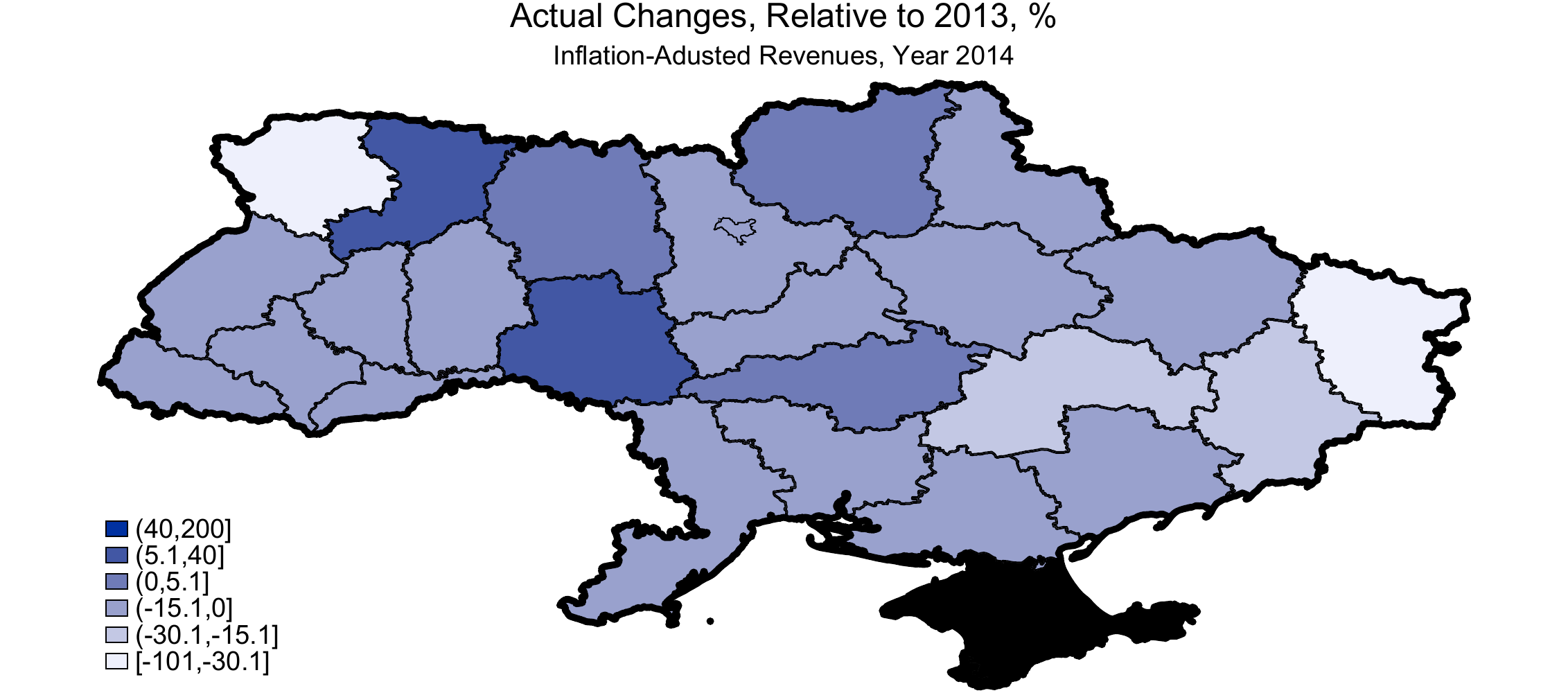} \\
        \includegraphics[width=1.0\textwidth]{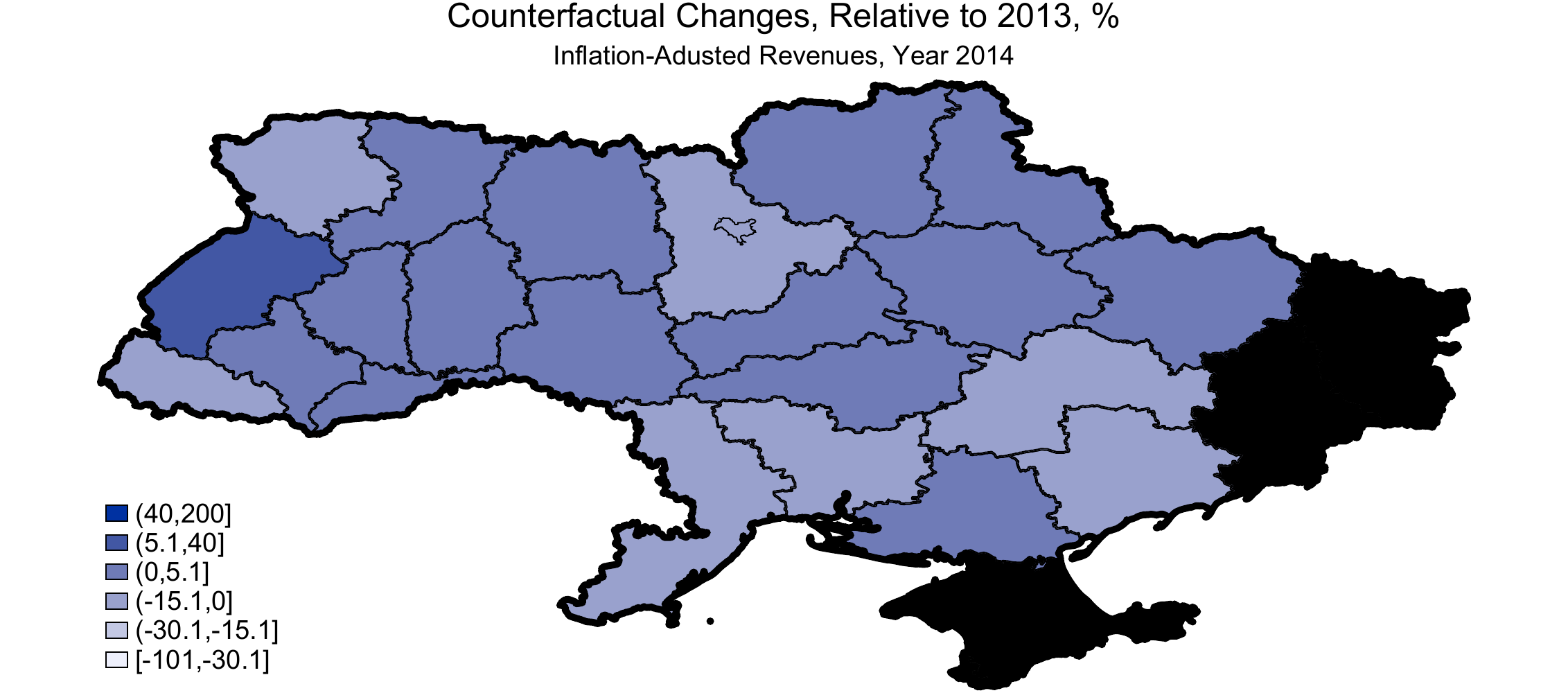}
    \end{center}
\caption{Network-Adjustment Counterfactual and Actual Changes, Province-Level}\label{FIG: NA Maps Province} %
\vspace*{0.01cm} 
\renewcommand{\baselinestretch}{0.7}
\footnotesize{\textit{Notes}: The graph depicts the evolution of raw and counterfactual revenues by Ukrainian provinces (oblasts). Darker colors indicate more positive changes. Black areas are either annexed or affected by the conflict. We use the black areas to estimate the outside demand $\bm{\xi}$, but not in reporting the counterfactuals. The top map displays the observed province-level changes of aggregate firm revenues in 2013--2014. The bottom map shows the changes in counterfactual revenues $\bm{r}$ for a given input-output matrix $\bold{\Omega}$ and outside demand $\bm{\xi}$ such that $\bm{r}(\bm{\xi},\bold{\Omega}):=[\bold{I}-(1-\alpha)\bold{\Omega}]^{-1}\bm{\xi}$, fixing outside demand at the 2013 level and using the actual 2014 railway trade matrix between provinces, $\bold{\Omega^{2014}_{province}}$.}
\end{minipage}
\end{center}
\end{figure}

\clearpage
\newpage

\begin{figure}[!t]
\begin{center}
\begin{minipage}{0.95\textwidth}
    \begin{center}
        \includegraphics[width=1.0\textwidth]{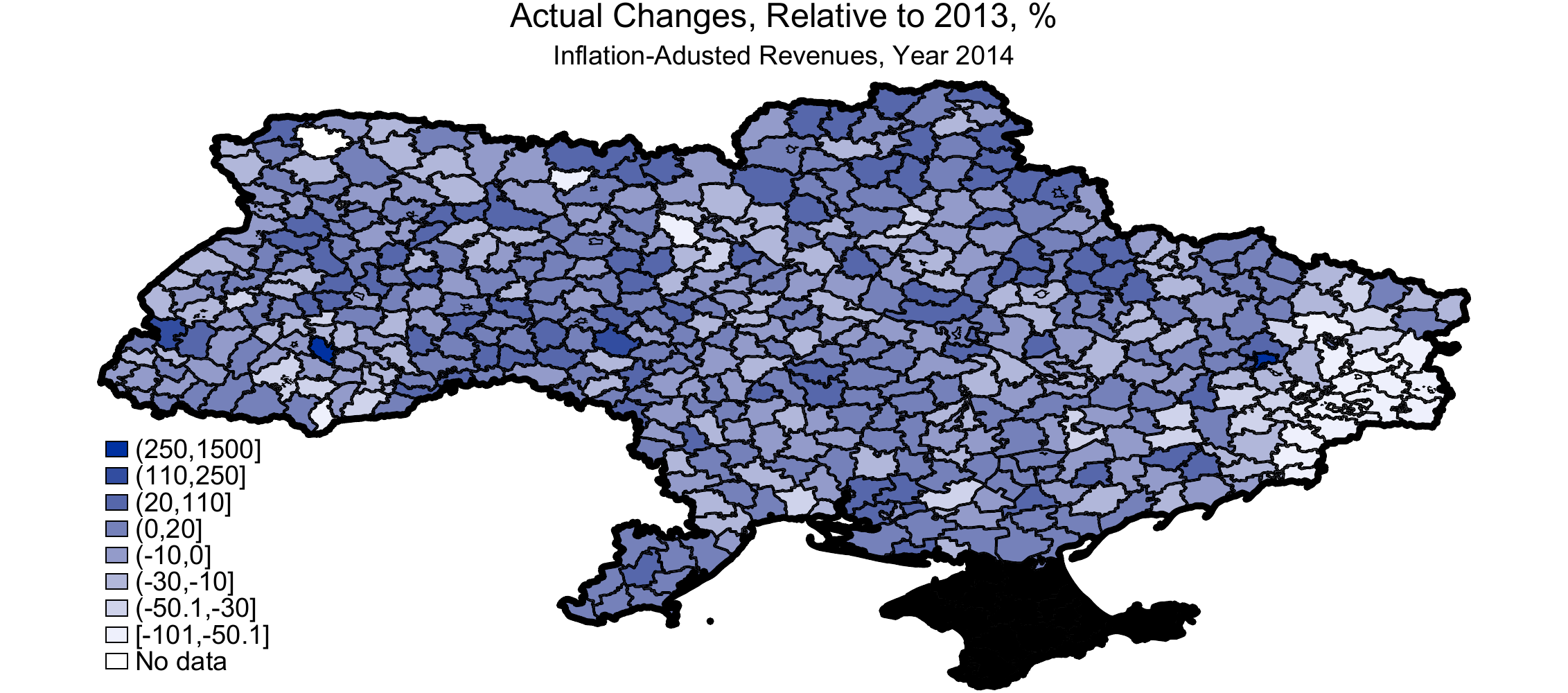} \\
        \includegraphics[width=1.0\textwidth]{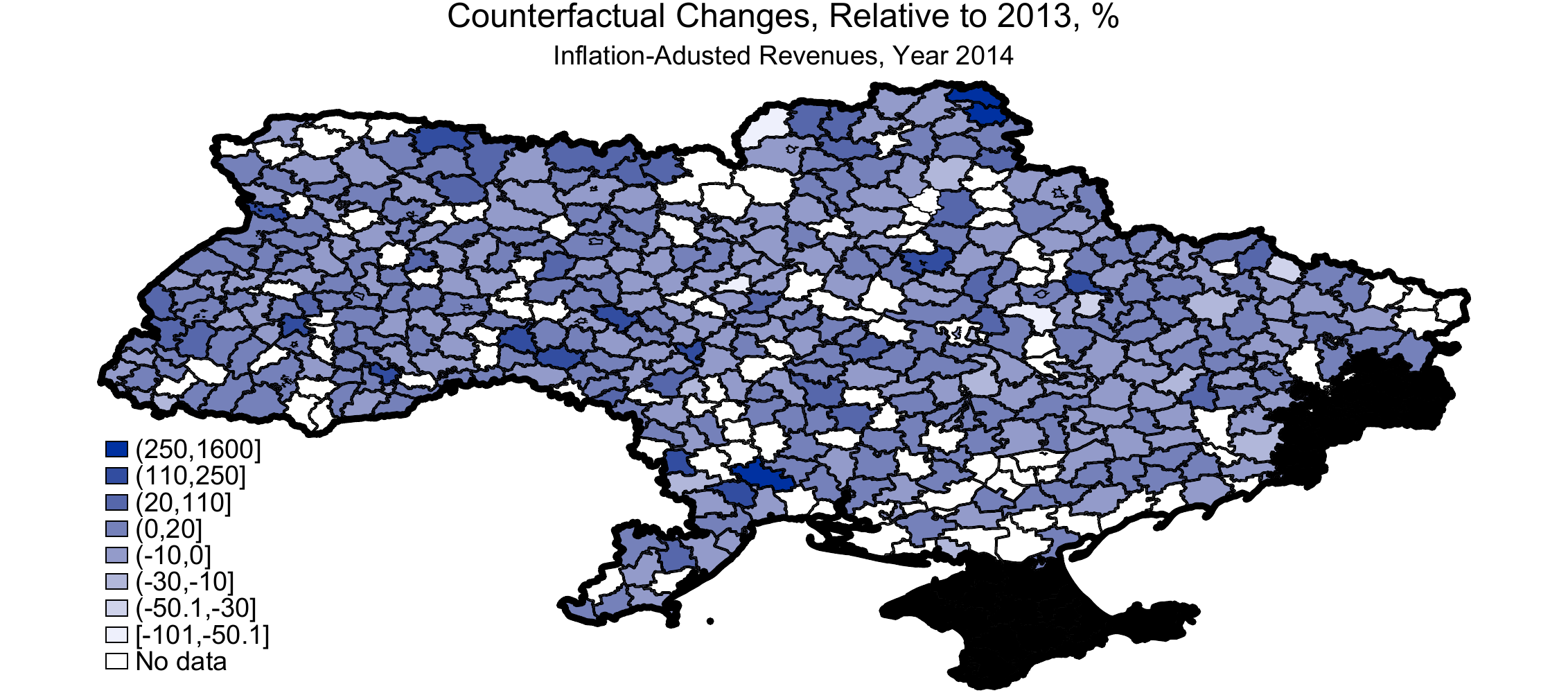}
    \end{center}
\caption{Network-Adjustment Counterfactual and Actual Changes, District-Level}\label{FIG: NA Maps District}%
\vspace*{0.01cm} 
\renewcommand{\baselinestretch}{0.7}
\footnotesize{\textit{Notes}: The graph depicts the evolution of raw and counterfactual revenues by Ukrainian districts (rayons). Darker colors indicate more positive changes. Black areas are either annexed or affected by the conflict. We use the black areas to estimate the outside demand $\bm{\xi}$, but not in reporting the counterfactuals. Missing data at the district level are depicted in white: those districts do not have railways. The top map displays the observed changes of aggregate firm revenues in 2013--2014. The bottom map shows the changes in counterfactual revenues $\bm{r}$ for a given input-output matrix $\bold{\Omega}$ and outside demand $\bm{\xi}$ such that $\bm{r}(\bm{\xi},\bold{\Omega}):=[\bold{I}-(1-\alpha)\bold{\Omega}]^{-1}\bm{\xi}$, fixing the outside demand at the 2013 level and using the actual 2014 railway trade matrix between districts, $\bold{\Omega^{2014}_{district}}$.}
\end{minipage}
\end{center}
\end{figure}

\end{small}
\end{spacing}

\clearpage
\newpage
\section*{Appendix B: Definitions of Centrality}\label{Appendix:Definitions}
\setcounter{equation}{0}
\setcounter{section}{0}
\setcounter{subsection}{0}
\setcounter{page}{1}
\renewcommand{\theequation}{\mbox{B\arabic{equation}}}
\renewcommand{\thesection}{\mbox{B\arabic{section}}}
\renewcommand{\thesubsection}{\mbox{B\arabic{subsection}}}
\renewcommand{\thetheorem}{\mbox{B\arabic{theorem}}}
\renewcommand{\theproposition}{\mbox{B\arabic{proposition}}}
\renewcommand{\thepage}{\mbox{B-\arabic{page}}}
\setcounter{table}{0}
\renewcommand{\thetable}{\mbox{B\arabic{table}}}
\setcounter{figure}{0}
\renewcommand{\thefigure}{\mbox{B\arabic{figure}}}

%\begin{spacing}{1.25}

To calculate degree centrality, we take an adjacency matrix $G = (g_{i,j})$, where $g_{i,j}$ is $1$ if firms $i$ and $j$ are linked and zero otherwise and define degree centrality for firms as a vector $C_d(G)$, where each entry is $C_{di} = \sum_{j=1}^N g_{ij}$. To define indegree and outdegree centrality, we replace $G$ with $G^o$ (and $G^i$), where $g_{i,j}$ is $1$ only if firm $i$ was selling (buying) to (from) firm $j$.

Betweenness centrality is a vector $C_b(G)$ where the $i$ entry is $C_{bi}=\sum_{k \neq j: i \{k,j\}}  P_i(k,j)/P(k,j)$, $P(k,j)$ is the total number of shortest paths from $k$ to $j$, and $P_i(k,j)$ is the number of those paths that pass through $i$ without $i$ being an endpoint.

We define eigenvector centrality, $C_e(G)$, from the matrix equation:
$$\lambda C_e(G) = G C_e(G),$$
where $\lambda$ is an eigenvalue corresponding to eigenvector $C_e(G)$. Following the consensus, we take the eigenvector associated with the largest eigenvalue.

%Finally, to calculate the weighted centrality measures, we replace $1$ in the definition of $G$ with the weights shipped.

%\end{spacing}
%%%

% \clearpage
% \newpage

\section*{Appendix C: Model Derivations}\label{Appendix:Model}
\setcounter{equation}{0}
\setcounter{section}{0}
\setcounter{subsection}{0}
\setcounter{page}{1}
\renewcommand{\theequation}{\mbox{C\arabic{equation}}}
\renewcommand{\thesection}{\mbox{C\arabic{section}}}
\renewcommand{\thesubsection}{\mbox{C\arabic{subsection}}}
\renewcommand{\thetheorem}{\mbox{C\arabic{theorem}}}
\renewcommand{\theproposition}{\mbox{C\arabic{proposition}}}
\renewcommand{\thepage}{\mbox{C-\arabic{page}}}
\setcounter{table}{0}
\renewcommand{\thetable}{\mbox{C\arabic{table}}}
\setcounter{figure}{0}
\renewcommand{\thefigure}{\mbox{C\arabic{figure}}}

\begin{spacing}{1.0}
We focus on the problem of the firm in this economy and derive the equilibrium conditions of interest from it:
\begin{gather*}
\max_{\{x_{ji}, l_i\}} \Big{\{} p_i x_i - wl_i - \sum_{j=1}^n x_{ji} p_j \Big{\}} , \text{ given }\\
x_i = (z_i l_i)^{\alpha} \Big{(} \sum_{j=1}^n a_{ji}^{1/\sigma}x_{ji}^{(\sigma-1)/\sigma} \Big{)}^{\frac{\sigma}{\sigma-1}(1-\alpha)}.
\end{gather*}
Deriving the first-order conditions for $l_i$ and $x_{ji}$ and multiplying by $l_i$ and $x_{ji}$, one gets the following system:
\begin{gather*}
\begin{cases}
     w l_i = \alpha p_i x_i,\\
    (1-\alpha) p_i x_i \cdot \frac{a_{ji}^{1/\sigma} x_{ji}^{\frac{\sigma-1}{\sigma}}}{\sum_{k=1}^n a_{ki}^{1/\sigma}x_{ki}^{\frac{\sigma-1}{\sigma}}} = p_j x_{ji}.
\end{cases}
\end{gather*}
Taking the ratio of the first-order conditions for two inputs $j$ and $m$ one gets
\begin{gather*}
    \frac{(a_{ji}/x_{ji})^{1/\sigma}}{(a_{mi}/x_{mi})^{1/\sigma}} = \frac{p_j}{p_m},
\end{gather*}
rearranging and solving for $x_{ji}$, one gets,
\begin{gather*}
 x_{ji} = \frac{a_{ji}}{p_j^{\sigma}} \frac{\sum_m p_m x_{mi}}{\sum_m p_m^{1-\sigma} a_{mi}}.
\end{gather*}
Next we use market clearing condition $x_i = \sum_{j=1}^n x_{ji} + c_i$ to derive further the equilibrium revenue of each firm:
$$r_i = p_i x_i = \sum_{j=1}^n p_i x_{ij} + p_i c_i$$
Then, expressing $x_{ij}$ from the first-order condition, and plugging it back into $r_i$, we get,
\begin{gather*}
  r_i =  \sum_{j=1}^n p_i \frac{(1-\alpha)p_j x_j }{p_i} \frac{a_{ij}^{1/\sigma} x_{ij}^{\frac{\sigma-1}{\sigma}}}{\sum_{k=1}^n a_{kj}^{1/\sigma}x_{kj}^{\frac{\sigma-1}{\sigma}}}  = \\
  (1-\alpha) \sum_{j=1}^n \underbrace{\frac{a_{ij}/p_i^{1-\sigma}(\mathbf{a})}{\sum_m a_{mj}/p_m^{1-\sigma}(\mathbf{a})}}_{\text{I-O weight }\omega_{ij}(\mathbf{a})}  \cdot r_j + \underbrace{p_i c_i}_{\text{outside demand } \bm{\xi}}.
\end{gather*}
\end{spacing}
%%%
\end{document}

%% file: Tables/Values.tex
\newcommand\underestimationPerc {67\%}

%% file: Tables/Table_1_Firm_Rayon_Month.tex
%                    &       %est1        &       %est2        &       %est3        &       %est4        &       %est5        &       %est6        \\[-0.1cm]
Post Crimea $\times$ &      -0.114$^{***}$&      -0.268$^{***}$&      -1.487$^{***}$&      -0.131$^{***}$&      -0.306$^{***}$&      -1.715$^{***}$\\[-0.1cm]
\textnormal{ } \textnormal{ }  Either Partner is in Conflict Area                   &     (0.006)        &     (0.013)        &     (0.072)        &     (0.006)        &     (0.015)        &     (0.078)         \\[0.1cm]
Post Crimea $\times$ &                    &                    &                    &      -0.025$^{***}$&      -0.054$^{***}$&      -0.326$^{***}$\\[-0.1cm]
\textnormal{ } \textnormal{ } Either Partner Traded with Conflict Areas                    &                    &                    &                    &     (0.003)        &     (0.008)        &     (0.038)        \\[0.1cm]
Establishment-Pair-Direction FE        &{\checkmark}        &{\checkmark}        &{\checkmark}        &{\checkmark}        &{\checkmark}        &{\checkmark}        \\[-0.1cm]
Year-Month FE       &{\checkmark}        &{\checkmark}        &{\checkmark}        &{\checkmark}        &{\checkmark}        &{\checkmark}        \\

Y Mean              &      0.077        &       0.181        &       1.005        &       0.077        &       0.181        &       1.005        \\[-0.1cm]
Y SD                &       0.267        &       0.724        &       3.509        &       0.267        &       0.724        &       3.509        \\[-0.1cm]
R$^2$               &        0.198        &       0.279        &       0.214        &       0.198        &       0.279        &       0.215        \\[-0.1cm]
Observations        &   11,001,648        &  11,001,648        &  11,001,648        &  11,001,648        &  11,001,648        &  11,001,648       \\[-0.1cm]
Province Pairs      &         751        &         751        &         751        &         751        &         751        &         751      \\

%% file: Tables/Table_1A.tex
                    &           .&            &            &            &            \\[-0.15cm]
Any Shipment        &  11,001,648&       .0771&        .267&           0&           1\\[-0.15cm]\addlinespace[0.15cm] 
Log Number of Shipments&  11,001,648&        .181&        .724&           0&          10\\[-0.15cm]\addlinespace[0.15cm] 
Log Total Weight Shipped&  11,001,648&           1&        3.51&           0&          22\\[-0.15cm]\addlinespace[0.15cm] 
Either Partner Is In Conflict Area&  11,001,648&        .105&        .307&           0&           1\\[-0.15cm]\addlinespace[0.15cm] 
Either Partner Traded with Conflict Area&  11,001,648&        .625&        .484&           0&           1\\[-0.15cm]\addlinespace[0.15cm] 
Supplier Is In Conflict Area&  11,001,648&       .0636&        .244&           0&           1\\[-0.15cm]\addlinespace[0.15cm] 
Customer Is In Conflict Area&  11,001,648&       .0415&        .199&           0&           1\\[-0.15cm]\addlinespace[0.15cm] 
Supplier Traded with Conflict Area&  11,001,648&         .48&          .5&           0&           1\\[-0.15cm]\addlinespace[0.15cm] 
Customer Traded with Conflict Area&  11,001,648&        .502&          .5&           0&           1\\[-0.15cm]\addlinespace[0.15cm] 
\addlinespace[0.15cm]

%% file: Tables/Table_1B.tex
                    &           .&            &            &            &            \\[-0.15cm]
Log of Firm Sales, 2010--2017&      27,187&       17.02&        2.42&        4.62&       27.25\\[-0.15cm]\addlinespace[0.15cm] 
IHS of Firm Profits, 2010--2017&      26,119&        6.93&       13.02&      -18.71&       24.79\\[-0.15cm]\addlinespace[0.15cm] 
Log of Firm Profits $-$ Log of Firm Total Costs, 2010--2017&      25,589&       -9.66&       12.60&      -42.27&       11.11\\[-0.15cm]\addlinespace[0.15cm]

%% file: Tables/Table_1_Firm_Month.tex
%                    &       %est1        &       %est2        &       %est3        &       %est4        &       %est5        &       %est6        \\[-0.1cm]
Post Crimea $\times$ &      -0.114$^{***}$&      -0.268$^{***}$&      -1.487$^{***}$&      -0.131$^{***}$&      -0.301$^{***}$&      -1.701$^{***}$\\[-0.1cm]
 \textnormal{ } \textnormal{ }  Either Firm Operated from Conflict Area                    &     (0.006)        &     (0.013)        &     (0.072)        &     (0.008)        &     (0.021)        &     (0.102)         \\[0.1cm]
Post Crimea $\times$ &                   &                    &                    &      -0.018$^{***}$&      -0.034$^{**}$ &      -0.225$^{***}$\\[-0.1cm]
\textnormal{ } \textnormal{ } Either Firm Traded with Conflict Areas                    &                     &                    &                    &     (0.005)        &     (0.015)        &     (0.068)       \\[0.1cm]

Establishment-Pair-Direction FE        &{\checkmark}        &{\checkmark}        &{\checkmark}        &{\checkmark}        &{\checkmark}        &{\checkmark}        \\[-0.1cm]
Year-Month FE       &{\checkmark}        &{\checkmark}        &{\checkmark}        &{\checkmark}        &{\checkmark}        &{\checkmark}        \\

Y Mean              &       0.077        &       0.181        &       1.005        &       0.077        &       0.181        &       1.005        \\[-0.1cm]
Y SD                &       0.267        &       0.724        &       3.509        &       0.267        &       0.724        &       3.509        \\[-0.1cm]
R$^2$               &       0.198        &       0.279        &       0.214        &       0.198        &       0.279        &       0.214        \\[-0.1cm]
Observations        &  11,001,648        &  11,001,648        &  11,001,648        &  11,001,648        &  11,001,648        &  11,001,648        \\[-0.1cm]
Province Pairs      &         751        &         751        &         751        &         751        &         751        &         751        \\

%% file: Tables/Table_1_Firm_Rayon_Month_BH.tex
%                    &       %est1        &       %est2        &       %est3        &       %est4        &       %est5        &       %est6        \\[-0.1cm]
Post Crimea $\times$ &      -0.115$^{***}$&      -0.271$^{***}$&      -1.505$^{***}$&      -0.127$^{***}$&      -0.298$^{***}$&      -1.663$^{***}$\\[-0.1cm]
\textnormal{ } \textnormal{ }  Either Partner is in Conflict Area                    &     (0.005)        &     (0.013)        &     (0.070)        &     (0.006)        &     (0.014)        &     (0.078)       \\[0.1cm]
Post Crimea $\times$ &                     &                    &                    &      -0.018$^{***}$&      -0.040$^{***}$&      -0.230$^{***}$\\[-0.1cm]
 \textnormal{ } \textnormal{ } Either Partner Traded with Conflict Areas                    &                    &                    &                    &     (0.002)        &     (0.005)        &     (0.025)        \\[0.1cm]
Post Crimea $\times$ &     -0.000$^{***}$&      -0.001$^{***}$&      -0.005$^{***}$&      -0.000$^{***}$&      -0.001$^{***}$&      -0.004$^{***}$\\[-0.1cm]
 \textnormal{ } \textnormal{ }  Number of Buyer's Partners in 2013                    &     (0.000)        &     (0.000)        &     (0.000)        &     (0.000)        &     (0.000)        &     (0.000)          \\[0.1cm]
Post Crimea $\times$ &      -0.000$^{***}$&      -0.000$^{*}$  &      -0.001$^{***}$&      -0.000        &      -0.000        &      -0.000\\[-0.1cm]
  \textnormal{ } \textnormal{ }  Number of Supplier's Partners in 2013                   &     (0.000)        &     (0.000)        &     (0.000)        &     (0.000)        &     (0.000)        &     (0.000)        \\[0.1cm]
Establishment-Pair-Direction FE        &{\checkmark}        &{\checkmark}        &{\checkmark}        &{\checkmark}        &{\checkmark}        &{\checkmark}        \\[-0.1cm]
Year-Month FE       &{\checkmark}        &{\checkmark}        &{\checkmark}        &{\checkmark}        &{\checkmark}        &{\checkmark}        \\

Y Mean              &       0.077        &       0.181        &       1.005        &       0.077        &       0.181        &       1.005        \\[-0.1cm]
Y SD                &       0.267        &       0.724        &       3.509        &       0.267        &       0.724        &       3.509        \\[-0.1cm]
R$^2$               &       0.198        &       0.279        &       0.215        &       0.199        &       0.279        &       0.215        \\[-0.1cm]
Observations        &  11,001,648        &  11,001,648        &  11,001,648        &  11,001,648        &  11,001,648        &  11,001,648        \\[-0.1cm]
Province Pairs      &         751        &         751        &         751        &         751        &         751        &         751        \\

%% file: Tables/Table_2_Firm_Rayon_Month_UD.tex
%                    &       %est1        &       %est2        &       %est3        &       %est4        &       %est5        &       %est6        \\[-0.1cm]
Post Crimea $\times$ &     -0.098$^{***}$&      -0.241$^{***}$&      -1.289$^{***}$&      -0.097$^{***}$&      -0.243$^{***}$&      -1.283$^{***}$\\[-0.1cm]
\textnormal{ } \textnormal{ } Supplier is in Conflict Area                    &     (0.007)        &     (0.017)        &     (0.093)        &     (0.008)        &     (0.017)        &     (0.097)        \\[0.1cm]
Post Crimea $\times$ &      -0.138$^{***}$&      -0.310$^{***}$&      -1.791$^{***}$&      -0.147$^{***}$&      -0.325$^{***}$&      -1.900$^{***}$\\[-0.1cm]
\textnormal{ } \textnormal{ } Buyer is in Conflict Area                    &     (0.007)        &     (0.018)        &     (0.088)        &     (0.007)        &     (0.019)        &     (0.090)        \\[0.1cm]
Post Crimea $\times$ &                    &                    &                    &      -0.011$^{***}$&      -0.027$^{***}$&      -0.148$^{***}$\\[-0.1cm]
\textnormal{ } \textnormal{ } Either Partner Had Supplier in Conflict Area                    &                    &                    &                    &     (0.002)        &     (0.005)        &     (0.027)         \\[0.1cm]
Post Crimea $\times$ &                    &                    &                    &      -0.023$^{***}$&      -0.043$^{***}$&      -0.294$^{***}$\\[-0.1cm]
\textnormal{ } \textnormal{ } Either Partner Had Buyer in Conflict Area                    &                     &                    &                    &     (0.003)        &     (0.008)        &     (0.039)        \\[0.1cm]
Establishment-Pair-Direction FE        &{\checkmark}        &{\checkmark}        &{\checkmark}        &{\checkmark}        &{\checkmark}        &{\checkmark}        \\[-0.1cm]
Year-Month FE       &{\checkmark}        &{\checkmark}        &{\checkmark}        &{\checkmark}        &{\checkmark}        &{\checkmark}        \\

Y Mean              &       0.077        &       0.181        &       1.005        &       0.077        &       0.181        &       1.005        \\[-0.1cm]
Y SD                &       0.267        &       0.724        &       3.509        &       0.267        &       0.724        &       3.509        \\[-0.1cm]
R$^2$               &       0.198        &       0.279        &       0.214        &       0.198        &       0.279        &       0.215        \\[-0.1cm]
Observations        &  11,001,648        &  11,001,648        &  11,001,648        &  11,001,648        &  11,001,648        &  11,001,648        \\[-0.1cm]
Province Pairs      &         751        &         751        &         751        &         751        &         751        &         751        \\

%% file: Tables/Table_1_Firm_Rayon_Month_Dist.tex
%                    &       %est1        &       %est2        &       %est3        &       %est4        &       %est5        &       %est6        \\[-0.1cm]
Post Crimea $\times$ &      -0.109$^{***}$&      -0.258$^{***}$&      -1.424$^{***}$&      -0.131$^{***}$&      -0.307$^{***}$&      -1.715$^{***}$\\[-0.1cm]
\textnormal{ } \textnormal{ }  Either Partner is in Conflict Area                    &     (0.005)        &     (0.012)        &     (0.067)        &     (0.005)        &     (0.013)        &     (0.067)        \\[0.1cm]
Post Crimea $\times$ &                    &                    &                    &      -0.030$^{***}$&      -0.065$^{***}$&      -0.386$^{***}$\\[-0.1cm]
\textnormal{ } \textnormal{ } Either Partner Traded with Conflict Areas                    &                    &                    &                    &     (0.002)        &     (0.005)        &     (0.024)        \\[0.1cm]
Post Crimea $\times$        &{\checkmark}        &{\checkmark}        &{\checkmark}        &{\checkmark}        &{\checkmark}        &{\checkmark}        \\[-0.1cm]
\textnormal{ } \textnormal{ } 5th Polynomial of Sender's Distance to Conflict Areas                   &                    &                    &                    &             &            &    \\[0.1cm]
Post Crimea $\times$         &{\checkmark}        &{\checkmark}        &{\checkmark}        &{\checkmark}        &{\checkmark}        &{\checkmark}        \\[-0.1cm]
\textnormal{ } \textnormal{ } 5th Polynomial of Receiver's Distance to Conflict Areas                   &                    &                    &                    &             &            &    \\[0.1cm]
Establishment-Pair-Direction FE        &{\checkmark}        &{\checkmark}        &{\checkmark}        &{\checkmark}        &{\checkmark}        &{\checkmark}        \\[-0.1cm]
Year-Month FE       &{\checkmark}        &{\checkmark}        &{\checkmark}        &{\checkmark}        &{\checkmark}        &{\checkmark}        \\

Y Mean              &       0.076        &       0.174        &       0.984        &       0.076        &       0.174        &       0.984        \\[-0.1cm]
Y SD                &       0.264        &       0.707        &       3.470        &       0.264        &       0.707        &       3.470        \\[-0.1cm]
R$^2$               &       0.195        &       0.278        &       0.212        &       0.196        &       0.278        &       0.213        \\[-0.1cm]
Observations        &   9,035,232        &   9,035,232        &   9,035,232        &   9,035,232        &   9,035,232        &   9,035,232        \\[-0.1cm]
Province Pairs       &         710        &         710        &         710        &         710        &         710        &         710        \\

%% file: Tables/Table_1_Firm_Rayon_Month_PostPrFE.tex
%                    &       %est1        &       %est2        &       %est3        &       %est4        &       %est5        &       %est6        \\[-0.1cm]
Post Crimea &      -0.077$^{***}$&      -0.177$^{***}$&      -1.000$^{***}$&      -0.103$^{***}$&      -0.233$^{***}$&      -1.332$^{***}$\\[-0.1cm]
\textnormal{ } \textnormal{ }  Either Partner is in Conflict Area &     (0.004)        &     (0.010)        &     (0.052)        &     (0.004)        &     (0.010)        &     (0.053)        \\[0.1cm]
Post Crimea &                    &                    &                    &      -0.029$^{***}$&      -0.064$^{***}$&      -0.379$^{***}$\\[-0.1cm]
\textnormal{ } \textnormal{ } Either Partner Traded with Conflict Areas                    &                    &                    &                    &     (0.002)        &     (0.004)        &     (0.022)        \\[0.1cm]
Establishment-Pair-Direction FE&{\checkmark}        &{\checkmark}        &{\checkmark}        &{\checkmark}        &{\checkmark}        &{\checkmark}        \\[-0.1cm]
Post Crimea-Province FE       &{\checkmark}        &{\checkmark}        &{\checkmark}        &{\checkmark}        &{\checkmark}        &{\checkmark}        \\[-0.1cm]
Year-Month FE       &{\checkmark}        &{\checkmark}        &{\checkmark}        &{\checkmark}        &{\checkmark}        &{\checkmark}        \\

Y Mean              &       0.077        &       0.181        &       1.005        &       0.077        &       0.181        &       1.005        \\[-0.1cm]
Y SD                &       0.267        &       0.724        &       3.509        &       0.267        &       0.724        &       3.509        \\[-0.1cm]
R$^2$               &       0.200        &       0.280        &       0.216        &       0.200        &       0.280        &       0.217        \\[-0.1cm]
Observations        &  11,001,648        &  11,001,648        &  11,001,648        &  11,001,648        &  11,001,648        &  11,001,648        \\[-0.1cm]
Province Pairs      &         751        &         751        &         751        &         751        &         751        &         751        \\